\documentclass[twocolumn]{aastex701}

\usepackage{amsmath}
\usepackage{natbib} 
\usepackage{graphicx}
\usepackage{xcolor} 
\usepackage{physics}
\usepackage{subcaption}
\usepackage{ulem}
\usepackage{booktabs}
\usepackage{multirow}

\def\W{\mathcal W}

\def\be{\begin{equation}}
\def\ee{\end{equation}}
\def\barr{\begin{array}{lr}}
\def\earr{\end{array}}
\def\bea{\begin{eqnarray}}
\def\eea{\end{eqnarray}}

\def\w{\omega}
\def\a{\alpha}
\def\b{\beta_s}
\def\c{\gamma}
\def\d{\delta}
\def\e{\epsilon}

\def\f{\phi}
\def\v{\varphi}
\def\g{\gamma}
\def\h{\eta}
\def\i{\iota}
\def\j{\psi}
\def\k{\kappa}
\def\l{\lambda}
\def\m{\mu}
\def\n{\nu}
\def\o{\omega}
\def\p{\pi}
\def\q{\theta}
\def\r{\rho}
\def\s{\sigma}
\def\t{\tau}
\def\u{\upsilon}
\def\x{\xi}
\def\z{\zeta}
\def\D{\Delta}
\def\F{\Phi}
\def\G{\Gamma}
\def\J{\Psi}
\def\L{\Lambda}
\def\O{\Omega}
\def\P{\Pi}
\def\Q{\Theta}
\def\S{\Sigma}
\def\U{\Upsilon}
\def\X{\Xi}

\begin{document}

\title{The \textsc{CatWISE2020} Quasar dipole: A Reassessment of the Cosmic Dipole Anomaly}

\author[orcid=0009-0002-2935-0496,sname='Bashir']{Masroor Bashir}
\affiliation{Indian Institute of Astrophysics, Koramangala II Block, Bangalore 560 034, India}
\affiliation{Department of Physics, Pondicherry University, R.V. Nagar, Kalapet, 605 014, Puducherry, India}
\email[show]{masroor.bashir@iiap.res.in}  

\author[orcid=0000-0002-7385-8273,gname=Pravabati, sname='Chingangbam']{Pravabati Chingangbam} 
\affiliation{Indian Institute of Astrophysics, Koramangala II Block, Bangalore 560 034, India}
\affiliation{Department of Physics, Pondicherry University, R.V. Nagar, Kalapet, 605 014, Puducherry, India}
\email{}

\author[0000-0001-8227-9516,sname=Appleby,gname=Stephen]{Stephen Appleby}
\affiliation{Asia Pacific Center for Theoretical Physics, Pohang, 37673, Republic of Korea}
\email{}

\begin{abstract}

The Ellis-Baldwin test probes the cosmological principle by comparing the kinematic Cosmic Microwave Background dipole with the Doppler-driven dipole in the number counts of extragalactic radio sources. Recent analysis of the CatWISE2020 quasar catalog reported a number-count dipole amplitude exceeding the kinematic expectation at $4.9\sigma$ significance. We present a comprehensive reassessment of this test using the same dataset, incorporating major sources of uncertainty in the statistical inference. We employ a simulation framework based on the \texttt{FLASK} package, using lognormal realizations of the large-scale structure, quasar clustering bias, the survey's radial selection function, and its exact sky coverage. Our simulations account for the kinematic dipole, the intrinsic clustering dipole, shot noise, and survey geometry effects. The analysis yields a revised significance of $3.63\sigma$ in the absence of a clustering dipole, and $3.44\sigma$ with a randomly oriented clustering dipole. When the clustering dipole is aligned with the kinematic dipole, the significance decreases further to $3.27\sigma$. Although the anomaly is reduced, it cannot be explained solely by the clustering dipole or mode coupling from the survey mask. We further assess dipole measurement robustness by fitting models with successively higher-order multipoles up to $\ell = 4$. Partial sky coverage induces mode coupling, shifting the dipole estimate to higher values when the octopole is included and inflating its variance as additional modes are incorporated, reflected in the increasing condition number of the estimator. This behavior highlights a bias-variance trade-off inherent in multipole fitting on partial-sky data.
\end{abstract}


\section{Introduction}

The cosmological principle, which proposes that the universe is statistically homogeneous and isotropic on sufficiently large scales, forms the bedrock of modern cosmology. This principle is directly tested through observations of the Cosmic Microwave Background (CMB) and large-scale structure (see e.g. \cite{Aluri:2023CQG}). The dipole anisotropy observed in the CMB temperature has been conventionally interpreted as a kinematic effect due to our  peculiar motion relative to the CMB rest frame. According to the most recent measurement the velocity inferred from CMB dipole measurements is $369.82 \pm 0.11~\text{km/s}$, in the direction $(l, b) = (264.^{\circ}021, 48.^{\circ}253)$ \citep{Planck_dipole:2020}.  

An important consequence of this kinematic interpretation is that it must be observable not only in the CMB but also as a corresponding Doppler-induced dipole anisotropy in flux-limited, all-sky surveys of sources. The  Ellis-Baldwin test can provide a crucial verification of this interpretation by searching for a Doppler-induced dipole in the number counts of distant radio sources \citep{Ellis:1984}. 
Given a population of sources with a homogeneous and isotropic  sky distribution, an observer's peculiar velocity induces a dipolar modulation in the flux and the number of sources per unit solid angle, from which a kinetic dipole amplitude can be inferred and compared with the value obtained from the CMB. 

With availability of a large number of sources in radio and infrared frequencies it became feasible to carry out the Ellis-Baldwin test.  
The 1.4 GHz NRAO VLA Sky Survey (NVSS) data \cite{WISE},  allowed astronomers to measure the dipole, providing an independent check of the kinematic interpretation 
\citep{Blake:2002, Singal:2011dy, Gibelyou:2012ri}. Subsequent analysis by \cite{Rubart:2013, Bengaly:2018bp, Singal:2019, Siewert:2021, Singal:2023, Wagenveld:2023, 
Land:2025, Bohme:2025} consistently reported dipole amplitude exceeding the CMB prediction, suggesting potential tensions with the standard kinematic interpretation.   

This mystery  deepened with more 
studies. 
\cite{Secrest:2021} measured a dipole in the \textsc{CatWISE2020} quasar catalog 
that was more than twice the CMB expectation at $4.9 \sigma$ significance level and aligned at $\sim$27$^\circ$ with respect to the CMB dipole direction. \cite{Dam:2023} confirmed this anomalously large signal using Bayesian inference. 
Further analysis of the same dataset was carried out by \cite{Tiwari:2023} testing the data against  \texttt{FLASK} simulations \citep{Xavier:2016} by incorporating quasar  clustering properties. They found a clustering dipole amplitude of $D_{\text{clus}} = 0.81 \times 10^{-3}$ ($C_1 = 0.91 \times 10^{-6}$). While substantial, this clustering component cannot fully account for the observed amplitude excess. Notably, for angular scales corresponding to multipoles $\ell \geq 10$ (approximately $>18^\circ$), the AGN/quasar clustering agrees with $\Lambda$CDM expectations. A newer catalog, presented in \cite{Secrest:2022}, extended the WISE selection to include $\sim 1.6$ million sources. Their analysis yielded a slightly lower but still anomalously high  dipole significance of $4.4\sigma$. A recent reassessment of the newer catalog by \cite{Hausegger:2025} reported a higher clustering dipole amplitude, $D_{clus} = 1.25 \times 10^{-3}$ ($C_{1} = 2.21 \times 10^{-6}$) suggesting that the clustering contribution in the extended catalog may be larger than previously inferred, though still insufficient to fully account for the observed dipole. 

More recent analysis by \cite{Abghari:2024} specifically demonstrate that the large sky cut ($\sim$52.6\%) in the \textsc{CatWISE2020} catalog induces strong mode coupling between low-order multipoles. Consequently, if low-order multipoles (in their example, $\ell = 3$) have amplitudes comparable to the dipole, this coupling can bias the inferred dipole amplitude.

This coupling can result in dipole amplitude errors large enough that consistency with the CMB dipole cannot be ruled out, highlighting the critical challenge of disentangling true cosmological signal from mask-induced systematics. The conclusions of \cite{Abghari:2024} were challenged in \cite{Secrest:2025wyu}, with questions raised about the magnitude of the higher $\ell$-mode amplitudes adopted.

 In this work, we perform a reassessment of the reported dipole anomaly by implementing a comprehensive treatment of uncertainties potentially underestimated in previous analyses discussed above. We choose to use the older \textsc{CatWISE2020} quasar catalog presented in \citet{Secrest:2021}, containing approximately $1.3$ million sources, for two main reasons. First, this enables a direct and consistent comparison with earlier studies most notably \citet{Secrest:2021}, which reports the highest dipole significance for a single quasar catalog ($\sim 4.9\sigma$) as well as with subsequent works that relied on the same dataset. Second, our primary objective is to study and quantify the impact of clustering-related systematics on the dipole amplitude and significance when applied to quasar number-count surveys on a partial sky. This is a broader methodological question, and our  conclusions extend beyond a specific catalog. Our analysis is facilitated by the fact that the clustering properties of this particular dataset have already been rigorously characterized by \cite{Tiwari:2023}, including its bias model, radial selection function, and intrinsic clustering dipole, all of which directly inform our simulations.

 We systematically quantify the impact of key systematic effects on the significance level of the \textsc{CatWISE2020} quasar dipole, including: shot noise contamination, higher-order clustering contributions, the amplitude and orientation of the intrinsic clustering dipole, mask-induced mode coupling, and the propagation of cosmic variance uncertainties in the clustering component. By incorporating these effects through simulations, we derive an updated statistical significance that reflects a more complete error budget, providing a updated evaluation of the dipole's consistency with the kinematic interpretation. 

 This paper is structured as follows. Section. \ref{sec:sec2} discusses  the details of the \textsc{CatWISE2020} quasar catalog and the sky mask geometry used for the analysis. 
 Section. \ref{sec:sec3} presents the formalism for the number count dipole and outlines the least-square estimator employed to measure the dipole amplitude. In section \ref{sec:s4}, we analyze some sources of uncertainties and bias that affect the dipole measurement. This includes a quantification of the bias introduced by mode mixing due to incomplete sky coverage and an assessment of the cosmic variance inherent to the clustering signal. Section \ref{sec:s5} describes our procedure for generating mock quasar catalogs, which are essential for statistical validation. 
 Section \ref{sec:sec6} presents our main results, where we quantify the statistical significance of the observed dipole under a series of distinct astrophysical scenarios. Finally, Section \ref{sec:sec7} provides a discussion of our findings and the implications of our results.

\section{\textsc{CatWISE2020} Quasar Catalog}
\label{sec:sec2}
The quasar sample used in this study is the catalog  constructed by \cite{Secrest:2021} from the \textsc{CatWISE} catalog \citep{Eisenhardt:2020, Marocco:2021}. \textsc{CatWISE} is derived from data from the \textit{WISE} all-sky satellite survey \citep{Wright:2010qw}, which imaged the sky in four infrared bands: $W_1$\,(3.4\,\textmu m), $W_2$\,(4.6\,\textmu m), $W_3$\,(12\,\textmu m), and $W_4$\,(22\,\textmu m). The \textsc{CatWISE} catalogue prioritizes the $W_1$ and $W_2$ bands due to their superior sensitivity and depth, being 95\% complete for sources brighter than $W_1 \leq 17.4$\,mag and $W_2 \leq 17.2$\,mag. 

The \cite{Secrest:2021} selection applies the mid-infrared colour cut $W1 - W2 \geq 0.8$ to identify AGNs. This criterion  identifies $61.9 \pm 5.4$\,AGNs\, per deg$^{2}$ to a depth of $W_2 \approx 15.0$\,mag, with a reliability of $\leq$95\% \citep{Stern:2012}. Deeper samples can achieve densities of $130 \pm 4$\,deg$^{-2}$ for $W_2 \leq 17.11$ with a decreased reliability of $\leq$90\% \citep{Assef:2013}. To ensure high-quality photometry for cosmological analysis, the sample also implements a strict magnitude cut of $9.0 < W1 < 16.4$ to remove bright, saturated sources and faint objects affected by uneven coverage. 

{\em Ecliptic bias correction}: 
The ecliptic bias is a systematic bias in the source count, resulting in a lower density of sources per pixel near the ecliptic poles and a lower density near the ecliptic equator. We correct for this artifact following \cite{Secrest:2021}, using their linear fit as a correction and weighting function in our calculations. 

\begin{figure}[h]
    \centering
    \includegraphics[scale=0.4]{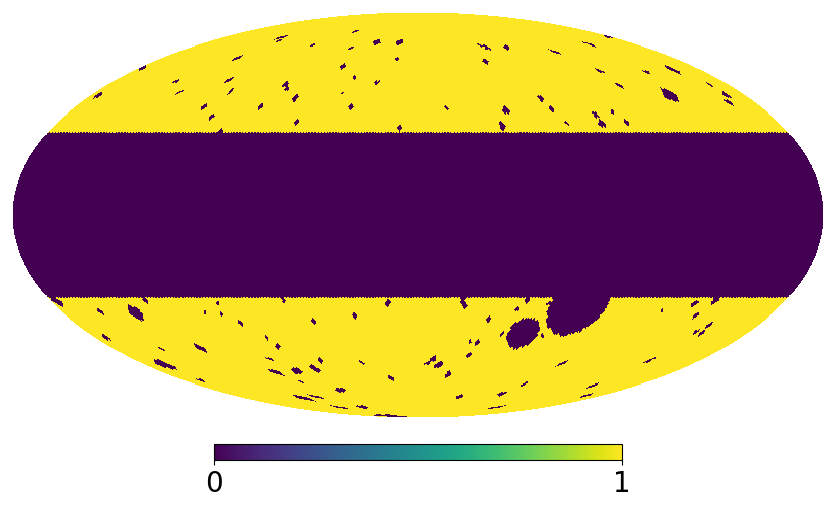}\\
        \includegraphics[scale=0.4]{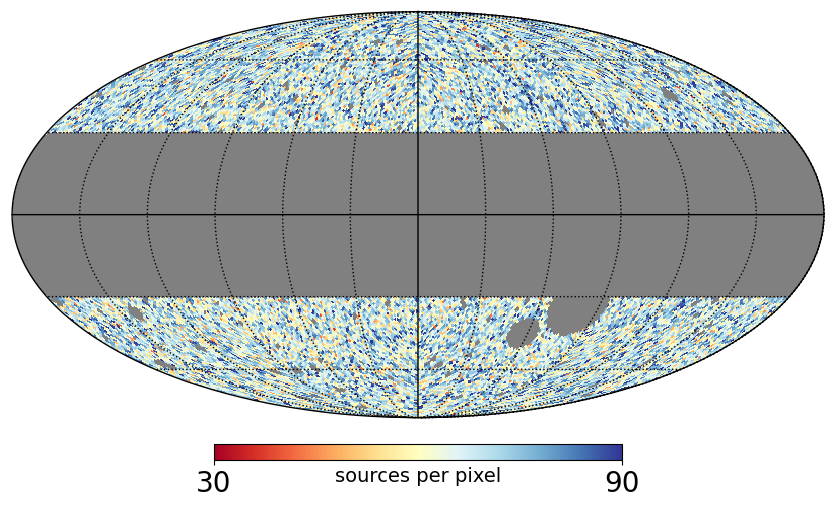}
    \caption{{\em Top}: Sky mask for the \textsc{CatWISE2020} quasar sample which excludes the Galactic plane ($|b| < 30^\circ$), regions around bright stars, and other areas with poor photometry, removing 52.6\% of the sky. {\em Bottom}: Sky distribution of the \textsc{CatWISE2020} quasar catalog used for dipole anisotropy studies.}
    \label{fig:fig1}
\end{figure}

{\em Sky mask}: For our analysis, we employ the \textsc{CatWISE2020} mask, depicted in the top panel of Figure~\ref{fig:fig1}. This mask is constructed by first applying a 
galactic cut ($|b| < 30^\circ$) which removes $50\%$ of the sky to avoid dust and stellar contamination from the Milky Way. Furthermore, 291 specific extragalactic point sources and extended objects are masked to eliminate contamination from local bright sources and nearby galaxies, removing an additional $\approx 2.6\%$ of the sky. The resulting total effective sky fraction is $f_{\text{sky}} \approx 0.474$ or 47.4\% sky coverage.

We compute the areal density of this sample by creating a map using the \textsc{HEALPix} framework \citep{Gorski:2005} with a resolution parameter $N_{\mathrm{side}} = 64$. The resulting masked density map is shown in the bottom panel of Figure \ref{fig:fig1}. The spectral properties of the sample, characterized by a mean spectral index of $\langle \alpha \rangle = 1.26$ and a mean logarithmic flux slope of $\langle x \rangle = 1.7$ are consistent with a predominantly AGN-dominated population.

\section{The number count dipole}
\label{sec:sec3}

Let $N(>S_{\nu},\hat n)$ denote the sky projected number count density of a given distribution of observed cosmological `sources' (e.g. CMB, number count densities of quasars, galaxies, X-ray sources, etc.), in sky direction $\hat n$.  $S_\nu$ denotes the minimum flux above which a source is detected.  $N(\hat n,>S_\nu)$ can be expanded as 
\begin{equation}
N(>S_\nu,\hat n) = \bar N\left( 1+ \mathbf{D}\cdot \hat n \right) + \mathcal{O}(\ell>1),
\label{eq:N}
\end{equation}
where $\bar N=N_{\rm tot}/4\pi f_{\rm sky}$ is the monopole term given by the average number density, with $N_{\rm tot}(>S_\nu)$ being the total number of observed sources with the survey area, and $\mathcal{O}(\ell>1)$ denotes terms  of higher multipoles higher than $\ell=1$.  
For the CMB and binned flux limited number counts averaged over a wide redshift range,  the higher multipole terms are usually small. The former because fluctuations in the early Universe are perturbatively small, and the latter because the act of smoothing large scale structure along the line of sight over gigaparsec scales washes out small scale clustering information. Hence for these particular datasets, the monopole and dipole are the dominant modes.  

The dipole vector, $\mathbf{D}$, represents the directional anisotropy in the angular distribution of cosmological sources. It is a vector sum of three 
distinct contributing vector quantities, 
\begin{equation}
\mathbf{D} = \mathbf{D}_{\text{kin}}  + \mathbf{D}_{\text{clus}} + \mathbf{D}_{\text{SN}}.
\label{eq:exactD}
\end{equation}
The physical origin of each component is distinct. The first one, $\mathbf{D}_{\text{kin}}$ is the \emph{kinematic dipole} which results from the observer's peculiar motion relative to an assumed cosmological rest frame. For a population of sources with a homogeneous and isotropic sky distribution and a flux density spectral index $\alpha$ (where $S_\nu \propto \nu^{-\alpha}$), a peculiar observer velocity $\vec{v}$ induces a dipolar modulation in the number of sources per unit solid angle, $N$, and in their flux densities, $S_\nu$. 
$N$ is commonly parameterized as a power law near the flux limit $S_\nu$ of a survey, as, $N(>S_\nu) \propto S_\nu^{-x}$. Then, the resulting amplitude of the dipole in $N$ caused by the Doppler boosting and aberration is given by (\citep{Ellis:1984}):
\begin{equation}
D_{\rm kin} = \big[2 + x(1+\alpha)\big]\beta,
\label{eq:Dkin}
\end{equation}
up to order $\beta$ in the limit $\beta\ll 1$, where $\beta=v/c$.


$\mathbf{D}_{\text{clus}}$ is the \emph{clustering dipole} which originates from large-scale matter inhomogeneities whose gravitational potentials modulate tracer distributions, generating an asymmetry that is intrinsic to the large-scale matter distribution. Lastly,  $\mathbf{D}_{\text{SN}}$,  is the \emph{shot-noise dipole} which is a statistical artifact stemming from finite-source Poisson sampling, scaling as $\sim 1/\sqrt{N}$ in magnitude. 

Expanding  $N(\hat n,S_{\nu})$ in spherical harmonics as $N(\hat n,S_{\nu}) = \sum_{\ell m} a_{\ell m}Y_{\ell m}$,  the power in each multipole mode is $C_{\ell} =\sum_m|a_{\ell m}|^2/(2\ell+1)$. 
For full sky data with $f_{\rm sky}=1$, we have $|\mathbf{D}|^2 \propto C_{1}$. 

For each dipole component $i$ (kin, clus or SN), 
we have $|\mathbf{D}_i|^2 \propto C_{1,i}$, where $C_{1,i}$ is the power of the $\ell=1$ component of the spherical harmonic expansion. 
The total dipole amplitude as derived by \cite{Gibelyou:2012ri} is written as,
\begin{equation}
|\mathbf{D}|^2   = \frac{9C_1}{4 \pi}, 
\end{equation}
where $C_1$ has contributions from all three components. 
For observed data, survey masks are employed to exclude regions contaminated by Galactic sources and other unwanted objects with $f_{\rm sky}<1$, breaking the sky's spherical symmetry. The asymmetry due to the survey mask 
leads to leakage of power from higher-order multipoles ($\ell \geq 2$) into the estimated dipole through mode mixing. Consequently, each dipole component becomes dependent on  multipoles higher than $\ell=1$, as $\mathbf{D}_i \equiv \mathbf{D}_i\left(C_{1,i},C_{2,i},\ldots \right)$. The dependence of each $\mathbf{D}_i$ on the multipole hierarchy is then determined by the mask geometry and the underlying power spectrum.

For a generic random field, $f(\hat n)$, we define its overdensity field as   
\begin{equation}
 \Delta f = \frac{(f(\hat{n}) - \bar{f})}{\bar{f}},   
 \label{eq:overdensity}
\end{equation}
where $\bar f$ denotes its mean value. In the sections that follow, we will use harmonic coefficients of both $f$ and $\Delta f$ depending on the context.

\subsection{Estimator for the number count dipole}
\label{sec:dipole_estimator}

For the sky pixelised into $N_{\rm pix}$ number of pixels, we use the following model for the number count density in each pixel indexed by $p=0,1,...,N_{\rm pix}-1$  (corresponding to sky direction $\hat n_p$)  given by (\cite{Secrest:2021}):
\begin{equation}
N(\hat n_p) = N_0 + \mathbf{D_1} \cdot \hat{n}_p + \epsilon_p.
\label{eq:dipole_est}
\end{equation}
$N_0$ is the parameter for the monopole $\bar N$, $\mathbf{D_1}=(D_{1,x}, D_{1,y}, D_{1,z})=\mathbf{D} N_0$ is the dipole vector in Cartesian coordinates. 
The last term $\epsilon_p$ encapsulates  contributions 
to the number count density $N(\hat{n}_p)$ that are not described by the simple monopole ($N_0$) and dipole ($\mathbf{D_1} \cdot \hat{n}_p$) components. These include higher order multipole contributions mentioned in Eq. \ref{eq:N}, and other sources of error. Masked pixels are excluded from the fitting procedure and assigned \texttt{NaN} values to prevent them from influencing Eq. \ref{eq:dipole_est} for unmasked sky.

We bin the source catalog to construct the random field $N(\hat n_p)$. From this we extract the monopole and the dipole by performing a linear least-squares fit between $N(\hat n_p)$  and the right hand side of Eq.~\ref{eq:dipole_est}, keeping upto the dipole term. The best-fit values for the monopole and dipole components, contained in the vector $\boldsymbol{\theta} = (N_0, D_{1,x}, D_{1,y}, D_{1,z})$, are found by minimizing the function:
\begin{equation}\label{eq:chi2}
\chi^2(\boldsymbol{\theta}) =
 \sum_{p}  w_p  \bigg[ N(\hat n_p) - \left( N_0 + \mathbf{D_1} \cdot \hat{n}_p \right) \bigg]^2,
\end{equation}
where the sum is over all unmasked pixels and $w_p$ are optional pixel weights (in this case $w_p=1$). We solve this system  using the \texttt{fit\_dipole} routine in the \textsc{HEALPy} package \citep{Zonca:2019}. 
The best-fit cartesian dipole vector is then normalized by the monopole to yield the dipole vector:
\begin{equation}
\mathbf{D} = \frac{1}{N_0} \begin{pmatrix} D_{1,x} \\ D_{1,y} \\ D_{1,z} \end{pmatrix}.
\end{equation}

Equivalently, the dipole amplitude $D=|\mathbf{D}|$ can be described using the angular power spectrum $C_\ell$ as (\citep{Abghari:2024}),

\begin{equation}
D = 3\sqrt{\frac{C_1}{C_0}}.
\end{equation}
This  connects the parameter-based fit ($N_0, \mathbf{D}$) to the harmonic-based power spectrum ($C_0, C_1$).

The fitting procedure outlined above, specifically minimizing (\ref{eq:chi2}), ignores the $\epsilon_p$ term in Eq.~\ref{eq:dipole_est}. Different physical and numerical effects will contribute to  $\epsilon_p$ which in turn add uncertainties in the measured values of $N_0$ and $\mathbf{D}_1$. 
While the simultaneous fit 
helps to isolate these components, it does not fully eliminate bias from higher-order moments.  These uncertainties will be discussed in the next section.

Throughout this work we make the distinction between `model', referring to the multipoles being fit to the quasar data, and `estimator' meaning the least squares fitting procedure as described in this section.   

\section{Uncertainties and bias in the estimation of the number count dipole}
\label{sec:s4}

In this section we examine how masking and cosmic variance broaden the uncertainties and potentially introduce bias in the measured dipole. Using the pseudo-$C_\ell$ mode-coupling matrix formalism for masked sky, developed in \cite{Hivon:2002}, we quantify the level of leakage of power across different modes for the \texttt{CatWISE2020} mask.  
Although a similar analysis was presented in \cite{Abghari:2024},  we  include it here for completeness of our analysis. Our goal is to clarify the underlying mechanism in a manner directly connected to the systematics investigated in this work. The insights gained here will be incorporated into the construction of realistic mock quasar number-count simulations in Section~\ref{sec:s5}.

\subsection{Mask induced effects - mode mixing}
\label{sec:sec4.1}

While masking removes major systematic contaminants, it simultaneously distorts the underlying signal by breaking the sky's isotropy and coupling power between different spherical harmonic modes. Consequently, the presence of higher-order multipoles such as those from large-scale structure clustering and residual systematic gradients can project into the dipole mode during the fitting procedure, thereby biasing the estimated dipole vector.


The observed harmonic coefficients $\tilde{a}_{\ell m}$ on the cut sky are a convolution of the true, full-sky harmonic coefficients $a_{\ell' m'}$ (of $N(\hat n,S_\nu)$). This relationship is encoded in the coupling kernel $K_{\ell m}^{\ell' m'}$ (\cite{Hivon:2002}):

\begin{equation}
\tilde{a}_{\ell m} = \sum_{\ell' m'} K_{\ell m}^{\ell' m'} a_{\ell' m'}.
\end{equation}
The effect of the mask on the angular power spectrum is characterized by the mode-coupling matrix $M_{\ell\ell'}$. The observed power spectrum $\langle \tilde{C}_{\ell} \rangle$ is a weighted sum of the true spectrum $\langle C_{\ell'} \rangle$:
\begin{equation}
\langle \tilde{C}_{\ell} \rangle = \sum_{\ell'} M_{\ell\ell'} \langle C_{\ell'} \rangle.
\end{equation}

\begin{figure}[h!]
    \centering
    \includegraphics[width=.91\linewidth]{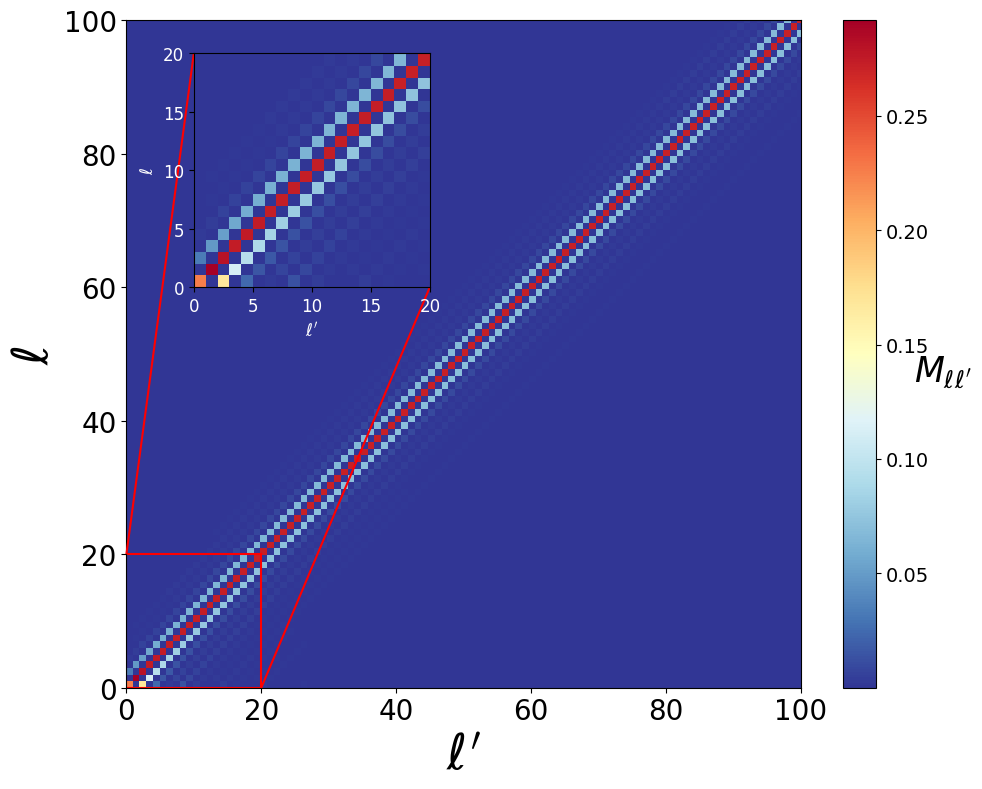}
    \includegraphics[width=.8\linewidth]{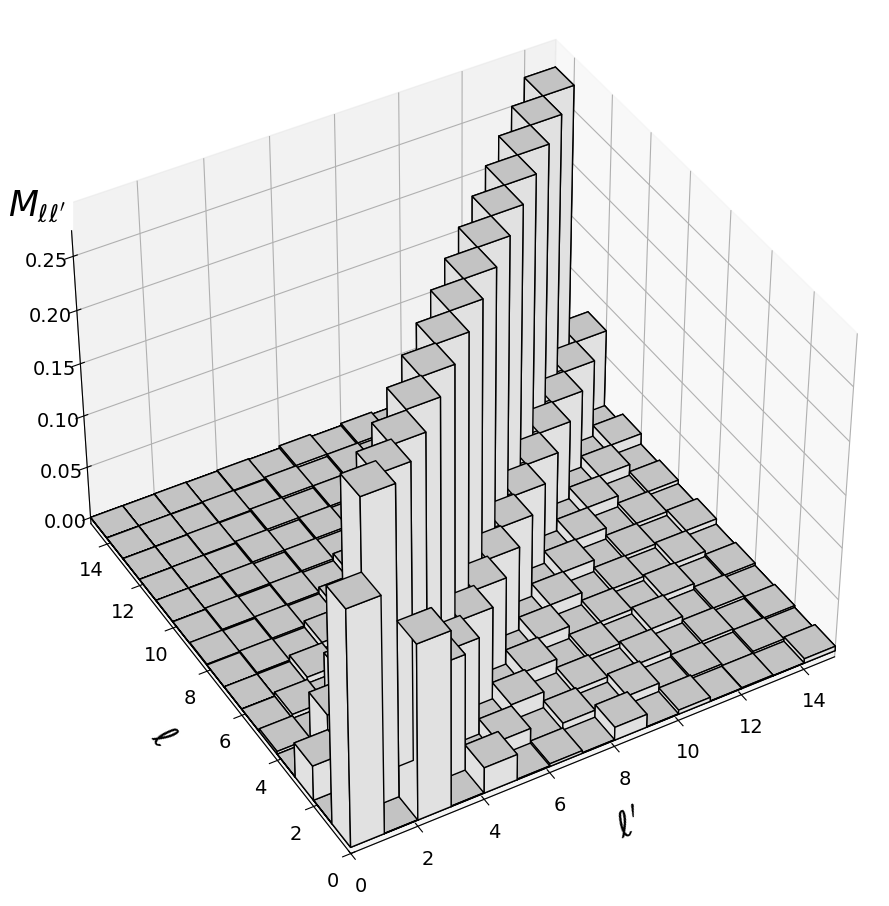}
    \includegraphics[width=.9\linewidth]{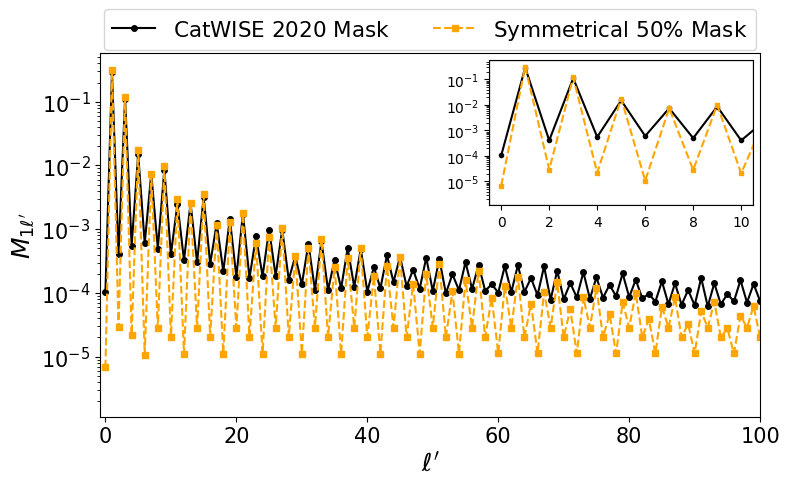}
    \caption{{\em Top and middle panels}: The mode-coupling matrix $M_{\ell\ell'}$ for the \textsc{CatWISE2020} survey mask 
    shown as a 2D heatmap (top) and as a 3D representation (middle). 
    Prominent off-diagonal elements demonstrate significant power leakage from higher multipoles ($\ell'$) into lower multipoles ($\ell$). 
    {\em Bottom panel}: $M_{1\ell'}$ plotted versus $\ell'$  illustrating the leakage of power from multipole $\ell'$ into the measured dipole ($\ell=1$). The plot clearly shows the decay of the coupling with increasing $\ell'$ and the enhanced leakage from odd multipoles compared to even ones (see the inset). For comparison, the coupling for a simple symmetrical galactic cut is also shown.}
    \label{fig:coupling_matrix}

\end{figure}

\noindent The mode-coupling matrix is determined by the power spectrum of the mask itself, as, 
\begin{equation}
M_{\ell\ell'} = \frac{2\ell' + 1}{4\pi} \sum_{L} (2L + 1) \mathcal{W}_L
\begin{pmatrix}
\ell & \ell' & L\\
0 & 0 & 0
\end{pmatrix}^2,
\end{equation}
where $\mathcal{W}_L$ is the angular power spectrum of the mask given by,
\begin{equation}
\mathcal{W}_L = \frac{1}{2L+1} \sum_{M} |w_{LM}|^2, 
\end{equation}
with $w_{LM}$ being the harmonic coefficients of the spherical harmonic expansion of the mask.

A major focus for this study are the matrix elements $M_{1\ell'}$, which quantify the leakage of power from higher-order multipoles $\ell' > 1$ into the measured dipole ($\ell=1$). Figure ~\ref{fig:coupling_matrix} demonstrates the nature of the leakage. 

These plots are similar in terms of information content to Fig. 6 of \cite{Abghari:2024}.
The top panel shows $M_{\ell\ell'}$ as a 2D heatmap. Prominent off-diagonal elements demonstrate significant power leakage from higher multipoles ($\ell'$) into lower multipoles ($\ell$). The middle panel provides a 3D representation of the same matrix, visually emphasizing the strength and pattern of this leakage. Both plots highlight how power from a wide range of scales couples with the dipole ($\ell = 1$), posing a significant challenge for an unbiased dipole measurement. 
    The behavior of $M_{1\ell'}$, plotted in the bottom panel  reveals two fundamental characteristics of this leakage for the \textsc{CatWISE2020} mask:
\begin{enumerate}
    \item \textbf{Decrease with $\ell'$:} The amplitude of $M_{1\ell'}$ diminishes with increasing $\ell'$, as the coupling to small-scale modes is low. 
    \item \textbf{Parity Effect:} The leakage is significantly stronger for odd-$\ell'$ multipoles compared to even ones. (See the inset in the bottom panel of Figure ~\ref{fig:coupling_matrix}.)
\end{enumerate}

Notably, when compared to an idealized symmetrical $50\%$ galactic cut, the more complex \textsc{CatWISE2020} mask with its additional small-scale masking produces a markedly higher level of leakage from even multipoles. Although this effect remains at the $\sim 1\%$ level for quadrupole ($\ell'=2$) and hexadecapole  ($\ell'=4$)  values considered here, the result highlights that even small changes to the mask geometry can affect the statistical significance of the recovered dipole, making an accurate characterization of the mask critically important.

To quantitatively understand the contamination of the dipole measurement by higher-order multipoles, we conducted a simulation analysis involving 1000 realizations of a sky map containing a dipole signal $\approx 0.007$ in the CMB dipole direction\footnote{We repeated our analysis with the dipole randomly directed on the sky, which did not alter our results.}. For each realization, we systematically injected a single higher multipole power - specifically the quadrupole ($C_2$), octopole ($C_3$), hexadecapole ($C_4$), or 
dotricontapole ($C_5$), with progressively increasing amplitude. The initial dipole amplitude $D_i$ (before injection) and the final amplitude $D_f$ (after injection) were measured after applying three distinct masks to the simulations: a $10^\circ$ symmetrical galactic cut, a $30^\circ$ symmetrical galactic cut, and the asymmetric \textsc{CatWISE2020} mask. For each mask, we computed the fractional deviation $(D_f - D_i)/D_i$ across all realizations, and analyzed its mean and $1\sigma$ dispersion as a function of the injected power.
This quantifies the bias induced by the mask in the recovered dipole amplitude.

\begingroup
\centering

\begin{figure*}[t!]
\centering
\hskip 1cm (a) \hskip 9cm (b)\\
\begin{subfigure}{0.49\textwidth}
\centering
\includegraphics[width=\linewidth]{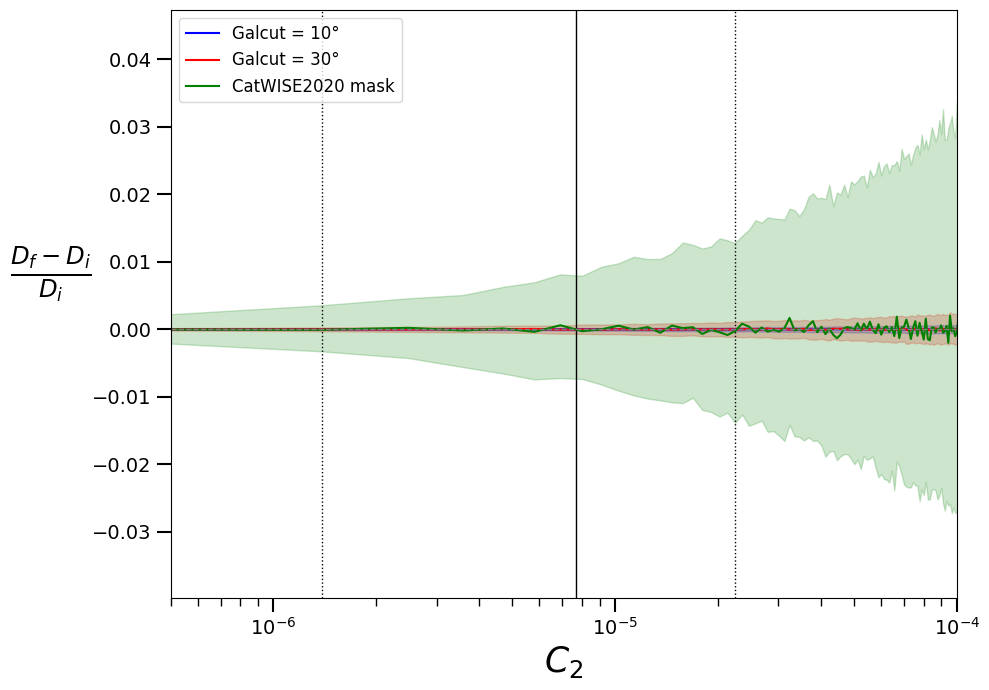}
\label{fig:variation_vs_c2}
\end{subfigure}
\hfill
\begin{subfigure}{0.49\textwidth}
\centering
\includegraphics[width=\linewidth]{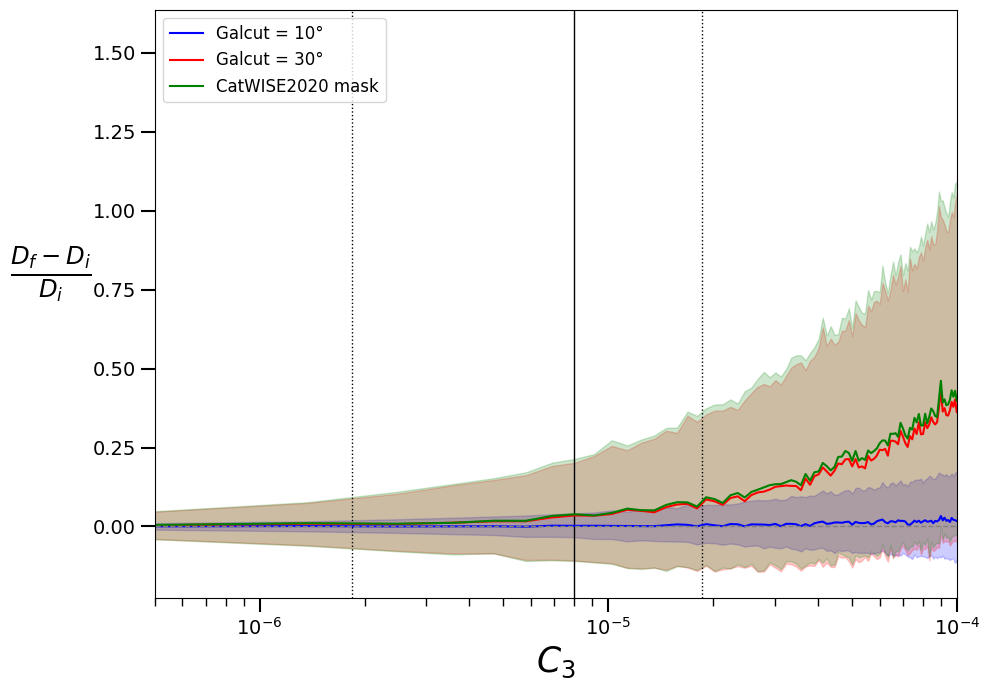}
\label{fig:variation_vs_c3}
\end{subfigure}
\vskip\baselineskip 
\hskip 1cm (c) \hskip 9cm (d)\\
\begin{subfigure}{0.49\textwidth}
\centering
\includegraphics[width=\linewidth]{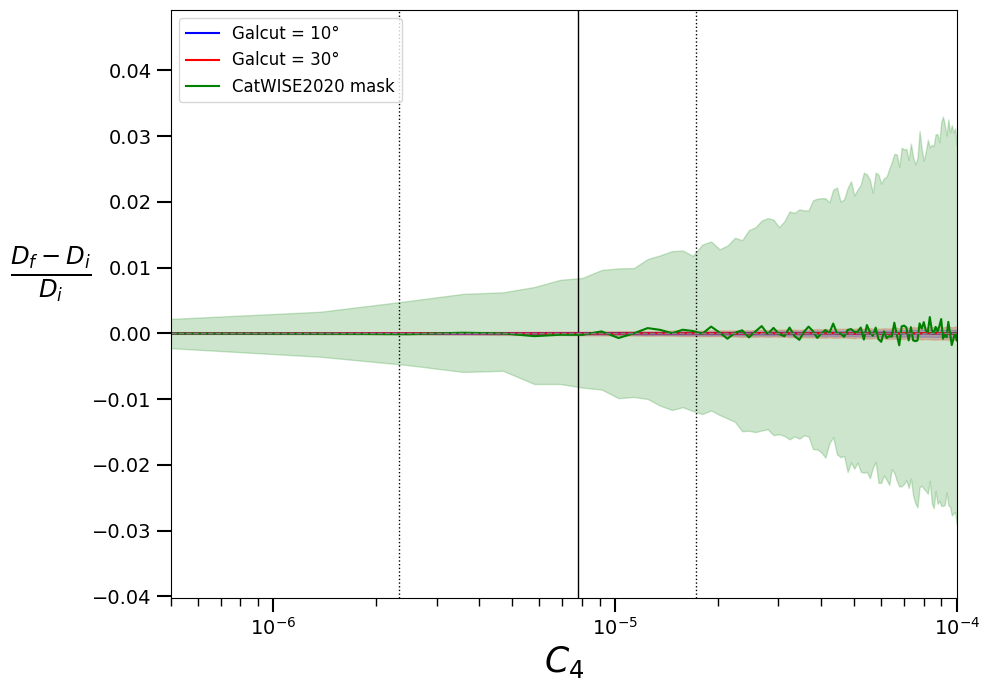}
\label{fig:variation_vs_c4}
\end{subfigure}
\hfill
\begin{subfigure}{0.49\textwidth}
\centering
\includegraphics[width=\linewidth]{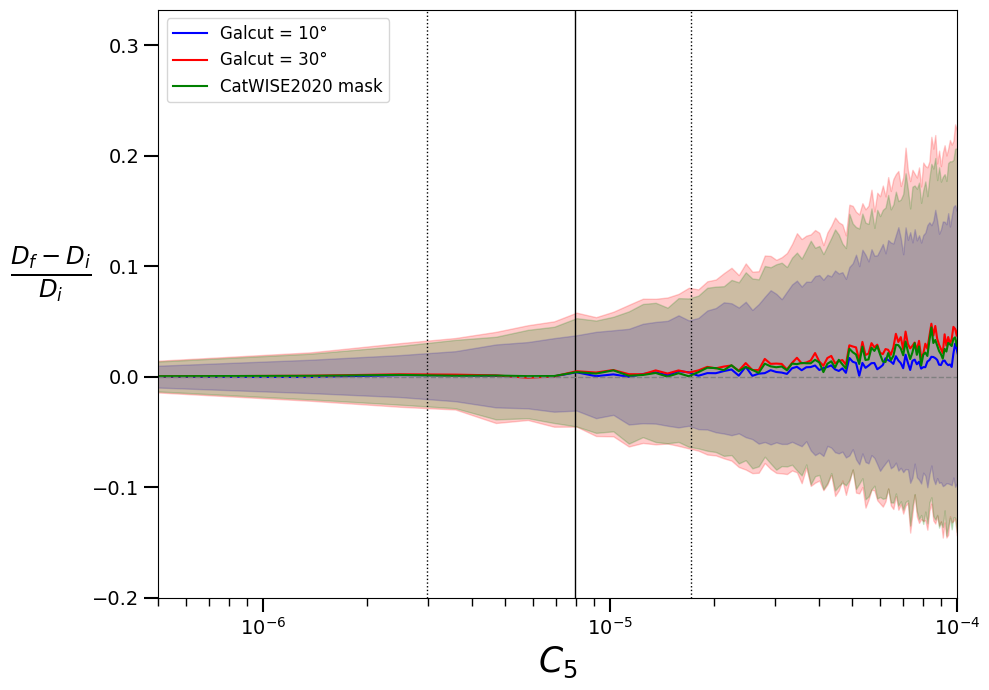}
\label{fig:variation_vs_c5}
\end{subfigure}
\caption{The fractional deviation of the measured dipole amplitude, $(D_f - D_i)/D_i$, as a function of injected multipole power. The solid lines represent the mean across 1000 realizations, while the shaded regions indicate the $\pm1\sigma$ uncertainty intervals. Results are shown for the \textsc{CatWISE2020} mask (orange) and symmetrical $10^\circ$ (green) and $30^\circ$ (blue) galactic cuts. Panels (a) and (c) show the results for even multipoles ($C_2$, $C_4$), where no systematic trend is observed. Panels (b) and (d) show the results for odd multipoles ($C_3$, $C_5$), demonstrating a clear positive correlation between the injected power and the fractional dipole deviation. For comparison, the solid and dotted black vertical lines represent the median $C_{\ell}$, and their corresponding 95\% confidence limits, 
from \textsc{FLASK} simulations. Note that the black vertical lines shows the \textit{full-sky} power spectrum ($C_\ell$) computed directly from the overdensity simulation maps via \texttt{healpy}.} 

\label{fig:variation}
\end{figure*}

\endgroup

The results for $(D_f - D_i)/D_i$, summarized in Figure \ref{fig:variation}, exhibit striking parity-dependent behaviour. For even multipoles, $C_2$, $C_4$, (panels (a) and (c)), the mean fractional deviation remains consistent with zero indicating no systematic bias. However, the scatter is more pronounced for the \textsc{CatWISE2020} mask, revealing  noise enhancement due to its asymmetry. In contrast, injections of odd multipoles ($C_3$, $C_5$) (panels (b) and (d)) produce a distinct positive, linear trend in $\langle (D_f - D_i)/D_i \rangle$, confirming a 
leakage of power that systematically inflates the dipole estimate and skews the underlying dipole probability density function towards larger values. Notably, the realistic \textsc{CatWISE2020} mask amplifies both effects,  yielding greater variance for even multipoles and bias  for odd ones. This underscores the critical role of mask geometry in modulating spurious mode coupling.

\subsection{Uncertainties due to  cosmic variance}
\label{sec:s4.2}

The angular kinematic dipole power, $C_{1, \text{kin}}$, is a deterministic  quantity for a particular observer, arising from the observer’s peculiar motion relative to the cosmic rest frame. 
In contrast, the clustering dipole, $C_{1, \text{clus}}$, is stochastic, originating from the particular distribution of matter in the  large-scale structure within our observable volume. As a result, its measured amplitude is inherently uncertain due to the scatter expected from observing one finite realization of a random field, or cosmic variance.  

The cosmic variance uncertainty for the clustering dipole is given by the formula:
\begin{equation}
\Delta C_{1,\text{clus}} 
=\sqrt{\frac{2}{3f_{\text{sky}}}} \, C_{1,\text{clus}}. 
\end{equation}
For $f_{\text{sky}} = 0.474$,   
the relative uncertainty at $1\sigma$ is approximately $118.6\%$ of $C_{1,\text{clus}}$.  
Consequently, the $1\sigma$ and $2\sigma$ intervals are:
\begin{align}
1\sigma &: \quad C_{1,\text{clus}} \pm 1.186 \, C_{1,\text{clus}}, \\
2\sigma &: \quad C_{1,\text{clus}} \pm 2.372 \, C_{1,\text{clus}} .
\end{align}
This large uncertainty propagates to the total observed dipole, significantly contributing to the error budget and complicating the extraction of the deterministic kinematic component. The exceptionally high cosmic variance for limited sky coverage underscores the challenge in achieving precise measurements of dipole anisotropies and highlights the need for careful modeling which takes these systematic effects into account.

\section{Mock Simulations}
\label{sec:s5}

The statistical significance of the dipole measured from an observed source catalog can be obtained by comparing with the dipole probability density function obtained from mock simulations. For accurate comparison, all physical effects that contribute to the dipole must be modeled and included in the simulations. 
An important shortcoming of the simulations used in \cite{Secrest:2021} is that they 
 included the kinematic dipole and shot-noise component, but not the clustering dipole (deeming it negligibly small) or correlated higher-order modes from large scale structures. 
By representing fluctuations only with a white-noise field of equal power across all multipoles, these simulations did not account for the bias that occurs from clustering power, especially from the odd multipoles,  that leaks into the measured dipole due to incomplete sky coverage. 
It was shown in \cite{Tiwari:2023} that the angular power spectrum of \textsc{CatWISE2020} for intermediate and high $\ell$ agrees well with $\Lambda$CDM expectation. This underscores the need to include these clustering properties in mock simulations. 

Here, we extend beyond previous works by improved modeling of the dipole contributions. We generate mock simulations of radio source catalogs 
incorporating the following: (a) kinematic dipole, (b) shot noise dipole systematics, (c) intrinsic higher $\ell$ large-scale structure and shot-noise contributions, and (d) clustering dipole, in this order.

(a) {\em Incorporating the kinematic dipole}: The kinematic dipole component, arising from our motion relative to the CMB rest frame, is simulated directly in pixel space for computational efficiency and physical clarity, following \cite{Secrest:2022}. For each pixel $i$ on the sphere, we compute the Doppler factor
\begin{align}
\delta_i = \frac{\left( 1 + \beta \cos \theta_i \right)}{\sqrt{1-\beta^2}},
\end{align}
where $\theta_i$ is the angular separation between the pixel center and the direction of the CMB dipole. $\beta$ is set to $1.23 \times 10^{-3}$, corresponding to the measured CMB dipole velocity of $v= 369.82$\,km/s in the direction $(l,b) \approx (264,~48)$ \cite{Planck_dipole:2020}.
The expected number of sources in each pixel is then given by
\begin{equation}
m_i = \delta_i^{\,2+x(1+\alpha)} \, N_0,
\label{eq:mi}
\end{equation}
which incorporates the combined effects of relativistic aberration and Doppler boosting, modifying the observed source distribution due to our peculiar motion relative to the CMB rest frame \citep{Ellis:1984}. Here, $N_0$ represents the sky-averaged monopole density, estimated from the total number count of sources in the survey. Note that the expression on the right hand side leads to the expression for $D_{\rm kin}$ given in Eq.~\ref{eq:Dkin}, upto order $\beta$. The values of parameters $x =1.7$ and $\alpha = 1.26$  are taken as the mean values derived from the \textsc{CatWISE2020} survey catalog, ensuring our simulations reflect the empirical characteristics of the observed quasar population.  This process produces a map dominated by the monopole and  dipole, with comparatively low power in higher multipoles ($\ell >1$). 

(b) {\em Incorporating shot noise dipole contributions}: To generate source counts, we populate each pixel by drawing a random number from a Poisson distribution with mean $m_i$, following \cite{Secrest:2022}. This process introduces the appropriate shot noise  corresponding to the discrete nature of source detection in flux-limited surveys. Shot noise contributes power across all multipoles, adding a scale-independent variance proportional to $1/\bar{N}$. This effect perturbs the initial monopole and dipole signal, introducing additional uncertainties in the measured dipole amplitude and direction, thereby increasing the error budget of the observed dipole. 

From the resulting maps, we extract the exact harmonic coefficients for the monopole ($a_{00}$) and the three components of the dipole ($a_{1,-1}, a_{1,0}, a_{1,+1}$) using a standard spherical harmonic transform. Here, we isolate only the monopole ($\ell=0$) and dipole ($\ell=1$) modes by setting all coefficients for $\ell \geq 2$ to zero, producing a map that contains purely the monopole, kinematic dipole signal and its associated shot noise dipole. This is done so as to not double count the $\ell\ge 2$ contributions of shot noise which are added in  step (c).

(c) {\em Incorporating higher multipole contributions}: We simulate higher-order multipoles ($\ell \geq 2$) using the \texttt{FLASK} package \cite{Xavier:2016}, which generates log-normal random fields  tomographically across 35 redshift bins from $z=0$ to $z=3.5$ ($\D z = 0.1$). The angular auto ($z_i = z_j$) and cross ($z_i \neq z_j $) matter power spectrum $C_\ell (z_i, z_j)$, which quantifies matter clustering across and within redshift bins, are computed using the Boltzmann code 
\texttt{CLASS} for a fiducial $\Lambda$CDM cosmology \cite{PlanckCP:2020}.
To connect the underlying matter distribution to the observable quasar clustering, we employ a redshift dependent best fit bias model $b(z) = 1.54 + 0.53z + 0.50z^2$ from \cite{Tiwari:2023} and the redshift distribution of the mean number of quasar sources denoted by $n(z)$, is modeled as $ n(z) \propto z^{a_1} \exp[-(z/a_2)^{a_3}]$, where the parameters $a_1 = 0.86, a_2 = 1.48$ and $a_3 = 2.36$ that define the shape, peak and tail of the distribution are adopted from \cite{Tiwari:2023}.

\begin{figure}[h!]
    \centering
    \vskip 0.6cm
    \includegraphics[width=.95\linewidth]{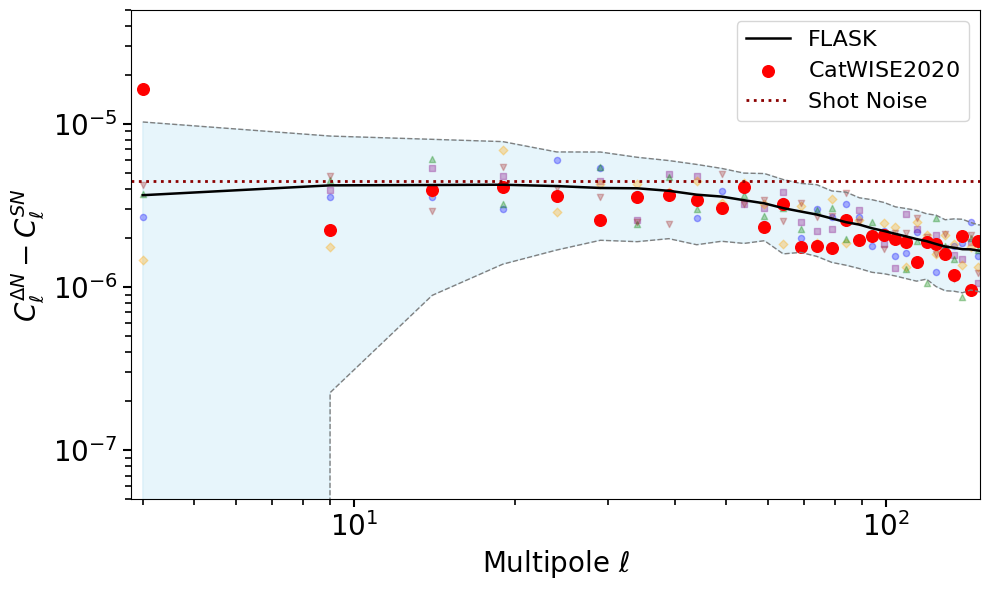}
    \caption{Shot noise-subtracted angular power spectrum ($C_\ell^{\Delta N}$ -  $C_\ell^{\rm SN}$) of the \textsc{CatWISE2020} quasar distribution overdensity field ($\Delta N$) 
    compared with masked \texttt{FLASK} simulations. The red points show the measured power spectrum from the real data, computed using NaMaster and averaged over 5 multipole bins. The solid black line represents the mean from 1000 \texttt{FLASK} mock realisations, while the shaded blue region indicates the $\pm 2 \sigma$ uncertainty derived from the simulations. The red dotted horizontal line represents the constant contribution from shot-noise ($C_{\ell}^{SN}$) over all modes. Faint colored symbols display individual power spectra from the first 5 \texttt{FLASK} realisations as generated in step (c) of section \ref{sec:s4}. The \textsc{CatWISE2020} measurements lie within the $2\sigma$ confidence interval for multipoles  $\ell > 5$, indicating consistency with the $\Lambda$CDM-based \texttt{FLASK} simulations on intermediate and small angular scales. This figure contains the same information as figure 6 of \cite{Tiwari:2023}. Note that the overdensity field includes a normalization of $1/\bar{N}^{\,2}$, producing the $\sim 4$-order-of-magnitude lower amplitude in Figure~4 relative to Figure~3. Shot-noise subtraction and masking further reduce the amplitude. Thus the spectra satisfy $C_{\ell}^{\Delta N} = C_{\ell}/\bar{N}^{\,2}$, where $C_{\ell}$ is the angular power spectrum of the \textsc{CatWISE2020} number density field, even though both reflect the same underlying clustering.}
    \label{fig:power_spectra}
\end{figure}

Let $C_\ell^{\rm SN}$ denote the angular power spectrum of shot noise, and  let $C_\ell^{\Delta N}$  denote the angular power spectrum of the overdensity field $\Delta N$ of \textsc{CatWISE2020} data (as defined in Eq.~\ref{eq:overdensity}). Figure \ref{fig:power_spectra} shows the shot-noise subtracted power spectrum $C_\ell^{\Delta N}$ -  $C_\ell^{\rm SN}$. The measured power spectrum agrees well with \texttt{FLASK} simulations within the 2$\sigma$ confidence interval for $\ell \geq 5$. This agreement validates that the amplitudes of higher multipoles considered in the simulations here are reasonable and representative of \textsc{CatWISE2020} quasar data. This comparison reveals  power in higher multipoles that was previously unaccounted for in dipole significance calculations, particularly the substantial clustering contribution at scales $3 \leq \ell < \ell_{\rm max}$ that can leak into the dipole measurement through mask effects. In general we use $\ell_{\rm max} = 3N_{\rm side}-1$ for our calculations. For the results that we will present in section~\ref{sec:sec6} we test the consistency of our analysis against variation of  $\ell_{\rm max}$ by systematically decreasing it. We find that $\ell_{\rm max} \simeq 64$ is sufficient to capture higher multipole effects, and increasing beyond this value does not significantly alter our results. 

\texttt{FLASK} first produces Gaussian random fields for each redshift shell, then applies a log-normal transformation to capture non-Gaussianity. Correlations between shells are preserved via the input power spectra. The line-of-sight integration across these shells yields a full-sky map, which is converted into a number count map by Poisson sampling, the number of quasars $N_q(i,j)$ in each pixel $i$ of the redshift shell $j$ is given by:
\begin{align}
N_q(i,j) = \bar{N_q}(j)[1 + \delta_q(i,j)] \, ,
\end{align}
where $\bar{N_q}(j)$ is the expected number of quasars per pixel in the redshift shell $j$ in a homogeneous universe and $\delta_q(i,j)$ is the quasar density field generated according to input power spectra. This introduces shot noise and discrete sampling effects. After generating the FLASK map, we compute its spherical harmonic coefficients and explicitly set the monopole ($\ell=0$) and dipole ($\ell=1$) modes to zero, retaining only coefficients for $\ell \geq 2$. This yields a map containing pure clustering (plus its shot noise) starting at $\ell=2$.

(d) {\em Incorporating the clustering dipole contribution}: The clustering dipole amplitude used in our simulations is $D_{\text{clus}} = 0.81 \times 10^{-3}$ ($C_1=0.91\times 10^{-6}$) adopted from \citet{Tiwari:2023}, who derived this value using their best-fit quasar redshift distribution ${n}(z)$ and bias evolution $b(z)$ mentioned in step (c) above. We generate this clustering dipole as an independent map containing only $\ell=1$ modes with the specified amplitude, subject to cosmic variance as characterized in section~\ref{sec:s4.2}.

Note that the value of $D_{\text{clus}}$ used here is higher than that determined by  \cite{Secrest:2021} ($D_{\text{clus}} = 0.24 \times 10^{-3}$). Their value is obtained with the assumption of a constant bias ($b(z) = 1$).

The final synthetic quasar number density map is a superposition of these carefully separated,  independent components. This construction ensures a clean separation of components with no double counting: the kinematic and shot-noise dipole originates exclusively from steps (a)+(b), the clustering dipole is added independently, and cosmological clustering and associated shot noise ($\ell\geq2$) comes solely from FLASK with its monopole and dipole modes explicitly set to zero.
\begin{widetext}
\begin{equation}
N(\hat{n}) = 
\underbrace{a_{00} Y_{00}(\hat{n})}_{\text{Monopole}~(N_{0})}\ + \hspace{-.3cm}\underbrace{\sum_{m=-1}^{1} a^{\rm kin}_{1m} Y_{1m}(\hat{n})}_{\text{CMB Kinematic Dipole}~(D_{\rm kin})}  \hspace{-.3cm} 
+ \hspace{-.1cm}\underbrace{\sum_{m=-1}^{1} a^{\rm SN}_{1m} Y_{1m}(\hat{n})}_{\rm SN\ Dipole~(D_{\rm SN})} 
\ + 
\underbrace{\sum_{m=-1}^{1} a^{\rm clus}_{1m} Y_{1m}(\hat{n})}_{\text{Clustering Dipole}~(D_{\rm clus})} 
+ \
\underbrace{\sum_{\ell=2}^{\ell_{\max}} \sum_{m=-\ell}^{\ell} a^{\rm SN+clus}_{\ell m} Y_{\ell m}(\hat{n})}_{\text{$\ell\ge 2$ Clustering + SN}}.
\label{eq:Nsim}
\end{equation}
\end{widetext}
 The different superscripts on $a_{\ell m}$s indicate the different contributions. To summarise, step (a) gives the first two terms, step (b) gives the third term, step (c) gives the fifth (last) term, and finally step (d) gives the fourth term. This equation captures the deterministic, signal-dominated kinematic dipole alongside the stochastic, log-normal fluctuations from large-scale structure and shot noise systematics. The fluctuations of the simulated $N(\hat{n})$ then provide a way to determine the statistical significance of the observed  \textsc{CatWISE2020} dipole amplitude. In figure  \ref{fig:f6} of appendix \ref{sec:a1},  we present one realization of each component, along with the composite $N(\hat{n})$.

In the next section, the statistical significance of the observed \textsc{CatWISE2020} dipole amplitude will be tested against different sets of simulations where we keep only specific contributions in Eq.~\ref{eq:Nsim}.  These are 
\begin{description}
\item[S0] Contains the monopole, CMB kinematic dipole, and a shot-noise component including all $1 \leq \ell \leq \ell_{\rm max}$ modes. These mocks reproduce the assumptions of \citet{Secrest:2021}.
\item[S1] Includes all contributions in Eq.~\ref{eq:Nsim} except the clustering dipole (the fourth term on the right-hand side). In particular, it adds higher-order  ($\ell \ge 2$) shot-noise and clustering moments (H.O.M) generated with \texttt{FLASK}, but omits the clustering dipole.  
\item[S2] Expands on {\bf S1} by adding a clustering dipole with amplitude taken from \citet{Tiwari:2023}. Its direction is chosen randomly on the sky, reflecting the assumption that large-scale structure is uncorrelated with our motion.
\item[S3] Identical to {\bf S2} except that the clustering and CMB kinematic dipoles are aligned,  representing the possibility of a common physical origin related to our local motion. The amplitude of each clustering dipole realization is varied in accordance with cosmic variance.The alignment is also motivated by findings such as those of \cite{Oayda:2024}, who observed that clustering dipole directions in data from the NRAO VLA Sky Survey and the Rapid ASKAP Continuum Survey  
appear to be aligned with the CMB dipole direction.
\item[S4] Similar to {\bf S2}, but with the clustering power ($C_{1,\rm clus}$) increased to its upper $2\sigma$ limit ($D_{\rm clus}^{2\sigma} = 1.48 \times 10^{-3}$); its direction remains random. This scenario is hypothetical, and tests the implications of a higher value of $D_{\rm clus}$ with random orientation.
\item[S5] In this scenario the clustering dipole direction is made aligned with the CMB kinematic dipole,  while the amplitude is constant and the same as in {\bf S4}. 
This  scenario is again hypothetical, and tests the implications of a higher value of $D_{\rm clus}$ with orientation aligned with the CMB.
\end{description}
These simulation sets are designed to isolate the impact of different physical inputs on the inferred statistical significance of the observed dipole amplitude. Specifically, {\bf S0} is used to verify consistency with \citet{Secrest:2021} under the same assumptions. {\bf S1}–{\bf S3} test the influence of the expected clustering dipole, which is respectively absent, randomly oriented, or aligned with the kinematic dipole in these three cases. Finally, {\bf S4} and {\bf S5} explore the sensitivity of the measured dipole significance to 
a clustering dipole with larger amplitude, that is randomly oriented or aligned with the kinematic dipole. 

\section{Results - Statistical significance}
\label{sec:sec6}

The observed \textsc{CatWISE2020} dipole amplitude, which we denote by $D^{\rm obs}$, is obtained to be  $15.54 \times 10^{-3}$, in agreement with the value reported by \cite{Secrest:2021}. This value was obtained using the standard monopole plus dipole $(M+D)$ 
model defined in section \ref{sec:dipole_estimator}.
To quantify the statistical significance of this observed value, we performed a hypothesis test against each suite of mock simulations prepared as described in section \ref{sec:s5}. We employed the \textsc{CatWISE} mask on the simulations to ensure  consistent sky coverage with the observed data set. Then we calculate the statistical significance for each set by counting how many simulations have dipole amplitude values that match or exceed $D^{\rm obs}$. The results are described below. 

\begin{figure*}
    \centering
    \includegraphics[width=1.0\linewidth]{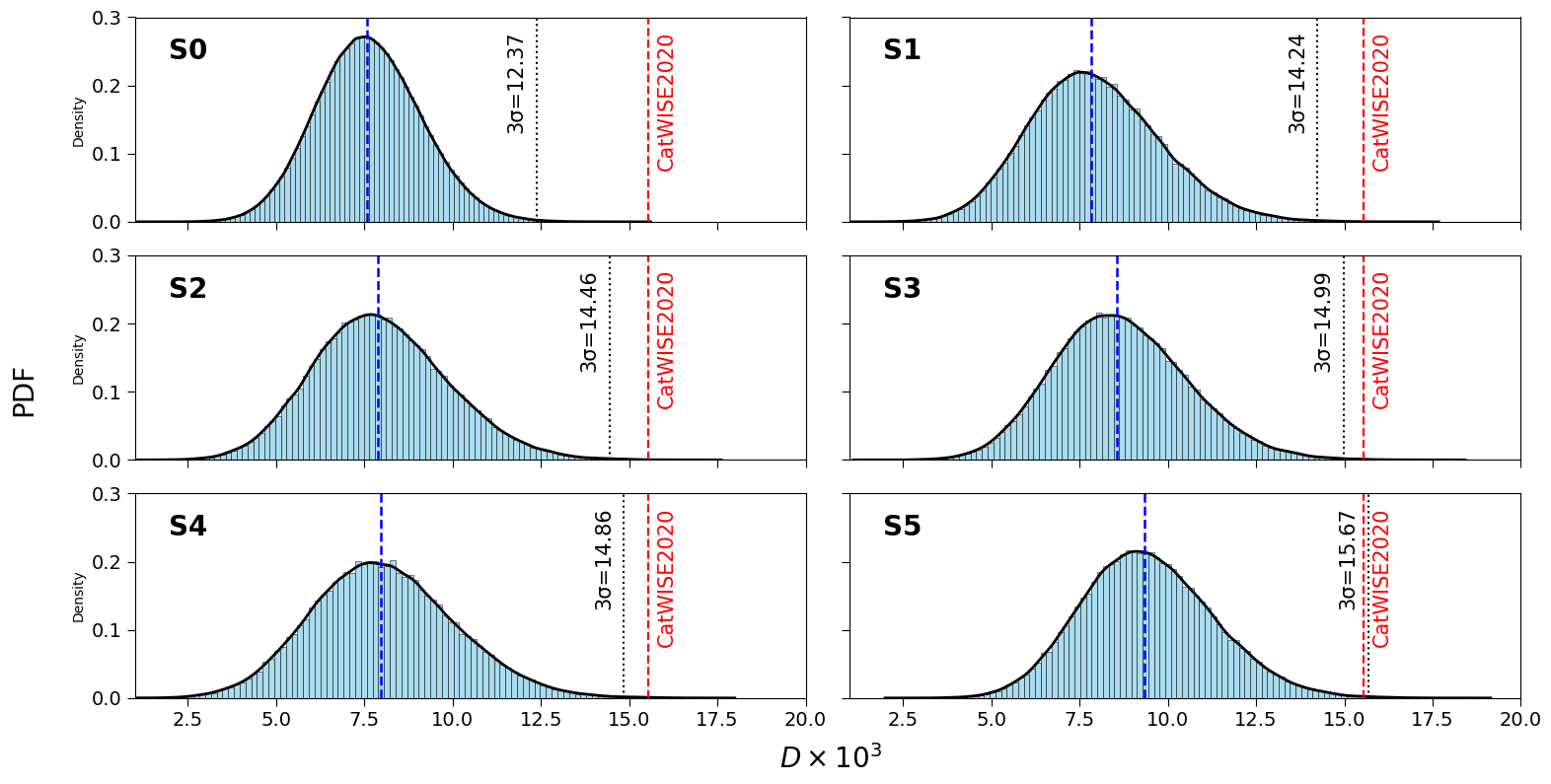}
    \caption{Probability density functions (PDFs) of the dipole amplitude ($\times 1000$) from six simulation scenarios. 
    Across all subplots, the \textsc{CatWISE2020} catalog dipole amplitude ($D^{obs}\times 10^3 = 15.54$) is indicated by the vertical red dashed line.  The blue dashed and black dotted vertical lines represent the median and  $3\sigma$ confidence level, respectively, specific to each simulation set.}
    \label{fig:f5}
\end{figure*}

{\bf S0 simulations}:  
This set has the same inputs as the simulations used by \cite{Secrest:2021} and serves to reproduce their results following their analysis pipeline. The top left panel of figure \ref{fig:f5} shows the PDF of the dipole amplitudes measured from the simulations. The \textsc{CatWISE2020}  value is marked in red.  We find 9 out of $10^7$ simulations have dipole values that matched or exceeded $D^{\rm obs}$ giving $p$-value $\approx 9 \times 10^{-7} $, or a significance of $4.82\sigma$. This confirms their reported high level of significance and verifies that our analysis is consistent with their study, before we introduce other physical effects in the simulations. 

From these simulations, we obtain  the ratio of the median values of $C_1$ and $C_3$ to be $\sim 18.01$. The $C_3$ power is contributed by shot noise alone, and it is roughly a factor of 18 smaller than $C_1$.


{\bf S1 simulations}: 
The top-right panel of Fig.~\ref{fig:f5} shows the PDF of the dipole amplitudes measured from the simulations. We observe that the PDF is broadened, accompanied by a slight shift of the median toward higher values relative to the {\bf S0} case. This behaviour can be attributed to the inclusion of stochastic clustering modes at multipoles $\ell \ge 2$, with the dominant contribution arising from $\ell = 3$, as discussed in Section~\ref{sec:s4}.

Among $10^5$ simulations, 13 exhibit dipole amplitudes equal to or exceeding the observed value $D^{\rm obs}$. This corresponds to a $p$-value of $1.30 \times 10^{-4}$, equivalent to a statistical significance of $3.63\sigma$.

The ratio of the median values of  $C_1$ and  $C_3$ is obtained to be approximately 8.71, which is roughly double of the value in the {\bf S0} case. The increase of the $C_3$ power results from  contributions from {\em both} shot noise and clustering. Nevertheless, $C_3$ remains approximately an order of magnitude smaller than $C_1$.

{\bf S2 simulations}: 
The left panel of the middle row of figure \ref{fig:f5} shows the PDF of the dipole amplitudes measured from the  simulations. We observe a mild shift of the median relative to the {\bf S0} case, similar to {\bf S1}. There is also a small broadening of the PDF due to the introduction of the stochastic clustering dipole relative to the PDFs of {\bf S0} and  {\bf S1} cases. 

We obtain 27 out of $10^5$ simulations matching or exceeding $D^{\rm obs}$. This yields a $p$-value of $2.70 \times 10^{-4}$, corresponding to a significance of $3.44\sigma$. 


{\bf S3 simulations}: 
The  right panel of the middle row of figure \ref{fig:f5} shows the PDF of the dipole amplitudes measured from these  simulations. We observe a shift of the median, relative to the previous cases, which can be attributed to the alignment between the clustering and kinetic dipole directions.

The number of simulations matching or exceeding $D^{\rm obs}$  increased to 58 out of $10^5$, corresponding to a $p$-value of $5.80 \times 10^{-4}$ and reducing the significance to $3.27\sigma$. 

{\bf S4 simulations}: 
The bottom left panel of figure \ref{fig:f5} shows the PDF of the dipole amplitudes measured from these simulations. There is a very mild broadening (not easily  discernible) of the PDF relative to {\bf S2}. This can be attributed to the enhanced clustering dipole amplitude compared to the value used in {\bf S2}.  

In this case, 53 out of $10^5$  simulations had dipole amplitudes that matched or exceeded $D^{\rm obs}$. This yields $p = 5.30 \times 10^{-4}$, corresponding to a significance of $3.27\sigma$.

 
{\bf S5 simulations}: The bottom-right panel of Fig.~\ref{fig:f5} shows the PDF of the dipole amplitudes measured from the {\bf S5} simulations. We observe a shift of the median relative to all previous cases, which can be attributed to the alignment between the clustering and kinematic dipole directions, as well as the enhanced amplitude of the clustering dipole. Among $10^5$ realizations, 175 yield dipole amplitudes equal to or exceeding the observed value $D^{\rm obs}$ ($p = 1.75 \times 10^{-3}$), corresponding to a significance of $2.93\sigma$. This indicates that the observed dipole remains a statistically significant anomaly even when all model parameters are pushed to their most extreme, yet still plausible, values.

\begin{table*}
\centering
\resizebox{\textwidth}{!}{%
\begin{tabular}{clcccc}
\hline
Simulations & Physical inputs & Clustering Dipole Direction & $n$ & $p$-value & Significance ($\sigma$) \\
\hline
{\bf S0} &  Kinematic dipole+Shot noise (Secrest et al. (2021))  & --- & 9 & $9.00 \times 10^{-7}$ & $4.82\sigma$\\   
{\bf S1} & Baseline + Higher Order Modes, No Clustering Dipole & --- & 13 & $1.30 \times 10^{-4}$ & $3.63\sigma$ \\
{\bf S2} & Baseline + Higher Order Modes + Clustering Dipole & Random & 27 & $2.70 \times 10^{-4}$ & $3.44\sigma$ \\
{\bf S3} & Baseline + Higher Order Modes + Clustering Dipole 
& Aligned with CMB Dipole & 58 & $5.80 \times 10^{-4}$ & $3.27\sigma$ \\

\hline
\end{tabular}%
}
\caption{Results of statistical significance tests from $10^5$ simulations for {\bf S1}-{\bf S3}, and $10^7$  realisations for the {\bf S0} case. We define `Baseline' here as the sum of the monopole ($N_{0}$), the kinematic dipole ($D_{kin}$), and the shot-noise dipole ($D_{SN}$) and higher order contributions; the first three terms on the right hand side of equation (\ref{eq:Nsim}). $n$ is the number of realisations that presented a dipole larger than the \textsc{CatWISE2020} value. The {\bf S0} case replicates the original Secrest et al. 2021 analysis, confirming their high significance.}
\label{tab:significance}
\end{table*}

In the five scenarios {\bf S1}–{\bf S5} analyzed here, the decrease in statistical significance relative to {\bf S0} arises from both the broadening of the PDFs and the shift of their medians toward higher values. This behaviour is naturally expected when the clustering dipole and higher-order multipoles are included, since their stochastic nature contributes positively to the total variance of the measured dipole. The extent of the broadening and median shift depends on the relative amplitudes of the dipole and higher multipoles. If the clustering dipole and higher multipoles were weaker than the values adopted in our analysis (\citealt{Tiwari:2023}), the inferred statistical significance of $D^{\rm obs}$ would be correspondingly higher.

The statistical significance values for the main physically motivated scenarios ({\bf S1, S2, S3}) are summarized in Table~\ref{tab:significance}, and constitute the main results of this paper. The value for the case of {\bf S0} is included for comparison. In conclusion, while the inclusion of a clustering dipole, particularly one that is aligned with the CMB dipole and affected by cosmic variance, increases the probability of observing a large dipole amplitude, the \textsc{CatWISE2020} result remains a statistically significant outlier at approximately the $3.3\sigma$ level or higher across all tested scenarios. 

To assess the robustness of the measurement of the observed dipole value, we further employed a suite of alternative 
models extending beyond the one defined in section \ref{sec:dipole_estimator}, sequentially including higher-order multipoles in the least squares fitting.  The full analysis, including discussions of their numerical stability, and the resulting dipole blue{values and their significance}, is presented in Appendix.~\ref{sec:a2}. In summary, while alternative 
models yield broadly consistent dipole directions, the inclusion of higher multipoles, particularly the octopole, increases the fitted dipole amplitude and markedly 
increases the variance of the 
measurement, as indicated by the condition number of the normal-equation matrix.

\section{Conclusion and Discussion}
\label{sec:sec7}

This work presents a reassessment of the cosmic dipole anomaly using the \textsc{CatWISE2020} quasar catalog. We have demonstrated the importance of accounting for several sources of uncertainties including a rigorous treatment of mode coupling introduced by the sky mask, which systematically biases the dipole estimate. 
Further, we developed mock simulations that accounts for kinematic, clustering and shot-noise effects. 

Our primary finding is a revised estimate of the statistical significance of the \textsc{CatWISE2020} dipole, summarized in Table~\ref{tab:significance}. The reduction from $4.9 \sigma$ to $3.63 \sigma$ in the case of no clustering dipole and to $3.27 \sigma$ in the case of CMB aligned clustering is a notable drop. However, it is clear that the observed, anomalously large dipole amplitude cannot be explained purely by the physical effects (clustering) and numerical artifacts (mode coupling) studied in this work, and it remains a  
challenge to the standard cosmological model. This conclusion is bolstered by similarly anomalous findings in other, independent data sets \citep{Colin:2017juj,Secrest:2022}.

Our analysis improves upon previous studies, most notably that of \cite{Secrest:2021} 
by incorporating a more complete treatment of systematic effects. While the earlier works correctly accounted for the shot-noise dipole,  
our study includes two missing components: the intrinsic clustering dipole and the influence of higher-order clustering modes. This inclusion is important. The sky mask induces mode coupling, but without the significant power from intrinsic clustering in the simulation, the previous significance estimates were large. 
Furthermore, we have systematically studied the bias introduced by the mask itself on the estimated dipole value and have employed \texttt{FLASK} simulation package to generate mock catalogs that replicates the survey's statistical properties, leading to improvement of the significance assessment. 
We also emphasize that if the amplitude of the clustering effects were lower than the values measured by \citet{Tiwari:2023} (which is used here), the inferred statistical significance of $D^{\rm obs}$ would increase, approaching the level reported by \citet{Secrest:2021}. 
Note that, for a direct comparison with the results of \cite{Secrest:2022}, we will need to repeat our analysis using their updated and expanded quasar catalog. We plan to undertake this as part of a follow-up study.

Finally, we systematically quantify how the statistical inference depends on the model itself, showing that  
including higher multipoles in the fitting increases the 
condition number and can obscure  the true level of tension. Crucially, we find that the choice of 
which low-order multipoles are simultaneously fitted, constitutes a significant systematic. High condition numbers 
that 
indicate mask-induced mode coupling can artificially suppress the measured significance by inflating the variance of the dipole PDF distribution, as described in Appendix. \ref{sec:a3}. 

\cite{Hausegger:2025} recently carried out a Bayesian inference based parametric fit and determined the dipole, quadrupole and octopole amplitudes 
using the extended catalog consisting of 1.6 million sources from \textsc{CatWISE2020}. The inferred 
dipole amplitude value varies slightly with the input model (only dipole, dipole+quadrupole, or dipole+quadrupole+octopole). 
They asked whether the anomalously large value of $D^{\rm obs}$ can be due to ignoring the octopole term, and find that including an octopole term slightly increases the observed dipole rather then decreasing it. They also state that the measured octopole value is consistent with noise based on a three sigma detection criteria. 
Our work differs from theirs in that our focus is the determination of the statistical significance of  $D^{\rm obs}$ (measured using Eq.~\ref{eq:chi2}) based on a frequentist hypothesis test when the clustering dipole and higher order clustering modes are included in mock simulations. 
Our results are broadly consistent with their analysis, the primary effect of including the octopole is to broaden the posteriors on the dipole, reducing its overall significance compared to the case in which only a dipole is fit to the data. 

A shortcoming of the present analysis is that contamination from stellar and other unwanted sources in the quasar catalog are not accounted for. Such  contamination can potentially alter the assessed significance level. The color cut of $W1 - W2 \geq 0.8$ produces a highly reliable ($\sim 95 \%$) quasar catalog \citep{Stern:2012}, but the impact of the remaining contaminants on the amplitude of \textsc{CatWISE2020} dipole is yet to be explored.  
In future work it is important to incorporate such contamination either by injecting into the simulations, or, utilize more refined observational datasets with robust unwanted source removal. Furthermore, a future study involving a morphological-level comparison between the \textsc{CatWISE2020} survey and our mock simulation can provide additional information. The large-scale behaviour of the one-point function of the data has been studied in \cite{Antony:2025tzk}, but more sophisticated techniques can be utilised to leverage more information. Such validation is essential for developing simulations  that accurately replicate the higher-order clustering statistics of the observed quasar population. 
Moreover, the least square estimator used here is susceptible to bias from mask-induced mode coupling and confusion with the intrinsic clustering dipole. Hence there is a need for improved 
methodologies  to disentangle these different effects~(\cite{Nadolny:2021}). 

Finally, estimations of the matter dipole have been extended to sources such as galaxies and quasars identified from optical observations. A tomographic dipole estimator applied to SDSS spectroscopic data found to broadly agree 
with the CMB dipole~\citep{Ferreira:2024,Tiwari:2024}.  
Similarly, a Bayesian analysis of the Quaia quasar sample \citep{Quaia:2024}, after applying conservative sky cuts, found the dipole consistent with the CMB \citep{Mittal:2024}, though this consistency may be driven by the larger noise in the reduced sample. 
\citet{Singal:2024ldf,Singal:2025} also analysed the Quaia data using a different selection function and concluded that the peculiar velocity of the Solar System is 4-5 times higher than the value inferred from the CMB dipole, with the direction pointing towards the Galactic centre within $1\sigma$. These results collectively highlight that conclusions drawn from matter dipole analyses remain sensitive to the choice of dataset and the methodology adopted, and the associated issues continue to be actively examined by the community \citep{Oayda:2025,Mittal:2025}. It is expected that forthcoming data releases with high source count densities such as from the Rubin telescope~\citep{LSSTScience:2009jmu,Braun:2019gdo} will enable a decisive measurement of the dipole, its redshift dependence, and the existence of the anomaly~ \citep{DESI:2016fyo,Amendola:2016saw,2020SPIE11443E..0IC}. 
Understanding the possible origin of such a dipole, and performing further tests of the statistical isotropy of large-scale structure, remain outstanding challenges in cosmology.


\begin{acknowledgments}
MB gratefully acknowledges Iqra Hamid Bhat for insightful discussions; Nidharssan S. and Vikranth P. for valuable suggestions; and Sriram K. and Arav B. J. for discussions on source systematics and mode mixing. The authors thank N. Secrest, S. V. Hausegger and S. Sarkar for critical comments on this work.  Computations were performed on the \texttt{Nova} cluster at the Indian Institute of Astrophysics, Bengaluru.  This work made use of \texttt{HEALPIX}\footnote{\url{http://healpix.sourceforge.net/}}~\cite{Gorski:2005}, \texttt{Healpy}~\cite{Zonca:2019}, \texttt{Matplotlib}~\cite{Hunter:2007}, \texttt{NaMaster}~\cite{Alonso:2018} and \texttt{CLASS}~\cite{Lesgourgues:2011}. 
The authors thank N. Secrest et al. for making their data publicly available. P.~C. and S.~A.  gratefully acknowledge the hospitality of Korea Institute for Advanced Study, Seoul, where part of this work was carried out.

\end{acknowledgments}
\appendix  
\section{Visualization of mock simulations}
\label{sec:a1}

\begin{figure*}
    \centering
  {\bf  Monopole \hskip 5.5cm Dipoles \hskip 3.5cm Higher multipoles $(\ell>1)$} \\
  \hskip 12cm ({\bf Shot noise + clustering})\\
  \vskip .2cm
    \hrule
    \vskip .2cm
    \begin{minipage}{0.29\textwidth}
    \centering
    \includegraphics[width=\linewidth]{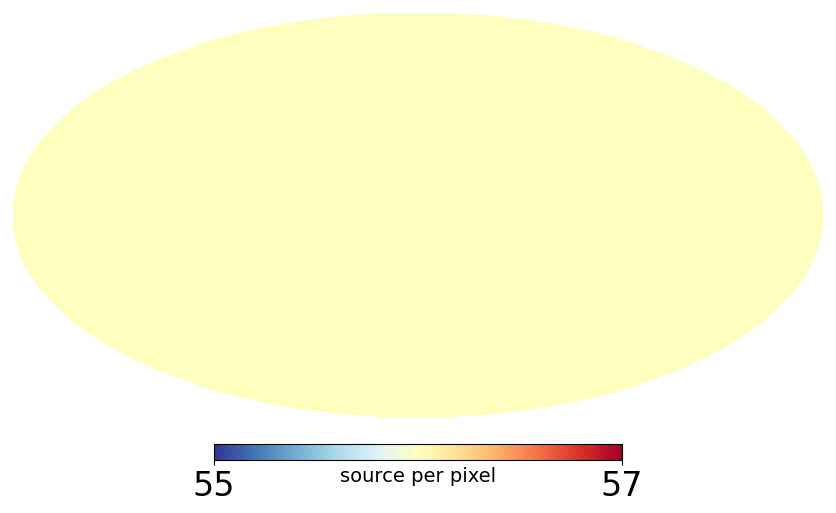}
    \end{minipage}%
    \hfill
    \begin{minipage}{0.29\textwidth}
        \centering
        {\bf Kinetic}\\
              \includegraphics[width=\linewidth]{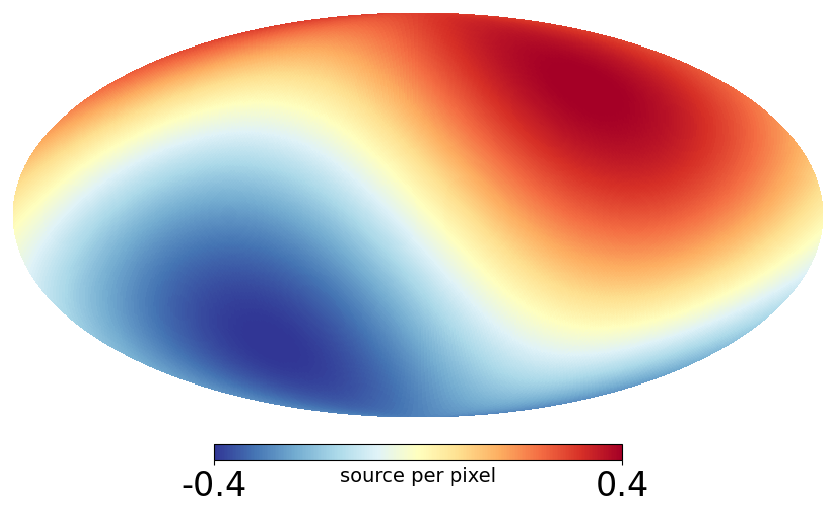}\\
        \vspace{2pt}
        {\bf Shot noise}\\
        \includegraphics[width=\linewidth]{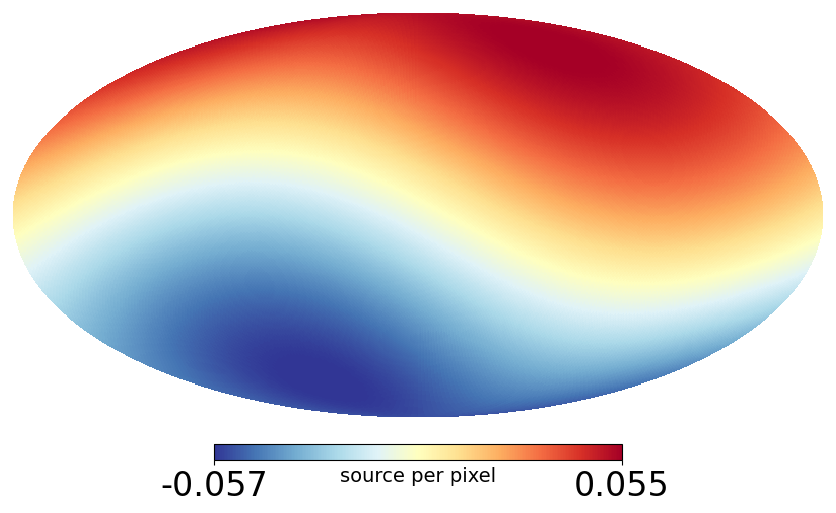}\\
        \vspace{2pt}
        {\bf Clustering}\\
        \includegraphics[width=\linewidth]{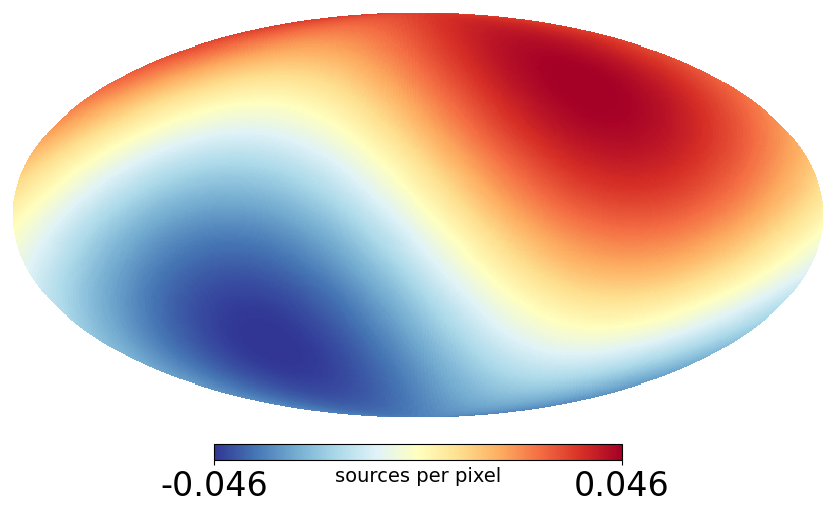}
    \end{minipage}
     \hfill
    \begin{minipage}{0.29\textwidth}
        \centering
        \includegraphics[width=\linewidth]{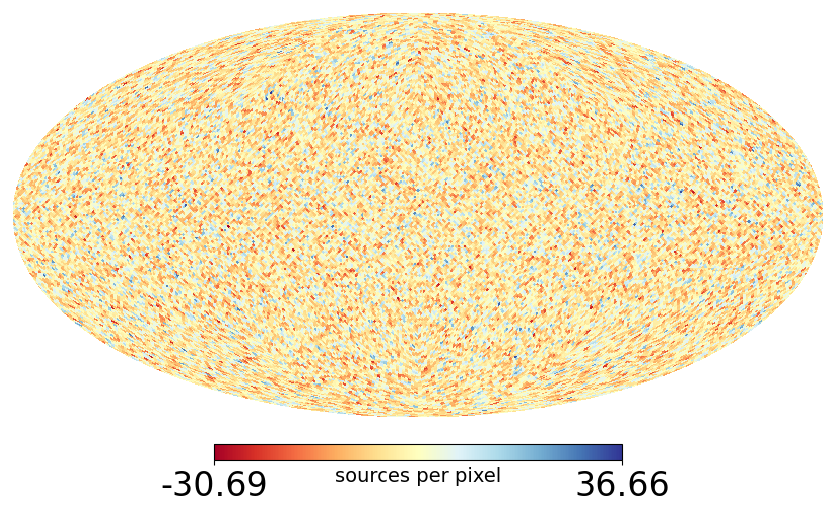}
    \end{minipage}%
    \vskip .5cm
{\Huge $\searrow \hskip 2cm \downarrow\hskip 2cm \swarrow$}
\vskip .5cm
    \begin{minipage}{0.3\textwidth}
        \centering
        \includegraphics[width=\linewidth]{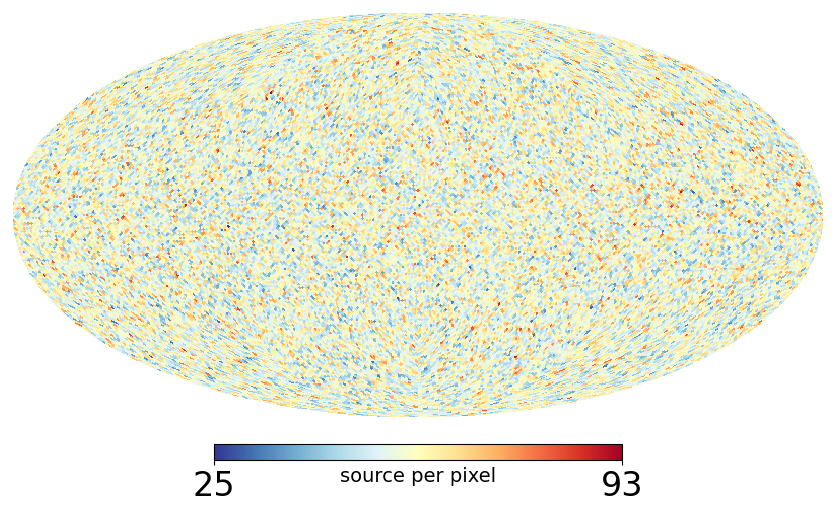}
    \end{minipage}
      \caption{{\em Left column}: Monopole component of the number count density map.
{\em Middle column}: Dipole contributions - kinematic (top), shot noise (middle), and clustering aligned with the CMB dipole direction (bottom).
{\em Right column}: Combined higher multipole contributions from shot noise and clustering. {\em Bottom}: Final simulated number count density map, $N(\hat n)$, as defined in Eq.~\ref{eq:Nsim}, obtained by summing all the above components. 
    }
    \label{fig:f6}
    \end{figure*}

For visual clarity, figure~\ref{fig:f6} illustrates the individual components that contribute to each simulation of $N(\hat n)$ (Eq.~\ref{eq:Nsim}). The left column shows the monopole component, which is uniform on the sky. The middle column displays the three dipole components: the CMB kinetic dipole (top), a shot noise dipole realization (middle), and a clustering dipole realization aligned with the CMB dipole direction (bottom). The right column shows the higher multipole contributions from shot noise and clustering. At the bottom, the composite number count density map, obtained by summing all components, is shown.

The complete maps $N(\hat n)$ include fluctuations arising from the shot noise and clustering dipoles, as well as higher-order multipoles. For full-sky data, only the kinetic, shot noise, and clustering dipole contributions would be relevant; however, masking induces mode coupling and makes the amplitudes of all fluctuations relevant for assessing the significance of the observed dipole in \textsc{CatWISE2020}.

\section{Estimation of Quasar dipole when including higher multipoles}
\label{sec:a2}

To estimate the direction and amplitude of the apparent dipole from the \textsc{CatWISE2020} quasar sample, we performed a series of sequential, low-multipole least-squares fits directly on the number density maps.

The spherical harmonic templates $\mathbf{Y}_{\ell m}$, masked to match the survey geometry, were used to form a linear model for the observed map:
\[
\mathbf{d}_{\text{model}} = \sum_{\ell \in S} a_{\ell m} \mathbf{Y}_{\ell m} = \mathbf{X} \mathbf{a},
\]
where $S$ is the set of multipoles included in the fit, $\mathbf{a}$ is the vector of corresponding spherical harmonic coefficients $a_{\ell m}$, and $\mathbf{X}$ is the design matrix whose columns are the mode vectors.

The coefficients $\mathbf{a}$ were obtained by minimizing the residual between the data vector $\mathbf{d}$ (the pixelated number density map) and the model:
\[
\min_{\mathbf{a}} \; \|\mathbf{d} - \mathbf{X}\mathbf{a}\|^2 .
\]
This leads to the normal equations
\[
\mathbf{X}^{\mathrm T}\mathbf{X}\,\mathbf{a} = \mathbf{X}^{\mathrm T}\mathbf{d},
\]
with the solution
\[
\mathbf{a} = (\mathbf{X}^{\mathrm T}\mathbf{X})^{-1} \mathbf{X}^{\mathrm T}\mathbf{d}.
\]

To monitor the stability of the inferred dipole against potential contamination from higher multipoles, particularly important given the partial sky coverage and the resulting mode coupling, we performed a sequence of fits with an increasing number of low multipoles. Specifically, we fit models containing: the monopole and dipole (M+D); the monopole, dipole, and quadrupole (M+D+Q); the monopole, dipole, quadrupole, and octupole (M+D+Q+O); and the monopole through $\ell=4$ (M+D+Q+O+$\ell=4$). For each fit, we compute the condition number\footnote{The condition number $C$ of a non-singular matrix $A$ is defined as $C=s_1/s_2$, where $s_1$ and $s_2$ are the largest and smallest singular values; large $C$ indicates numerical ill-conditioning.} \cite{belsley:1980} of the Gram matrix that is inverted to infer ${\bf a} : \kappa(\mathbf{X}^{\mathrm T}\mathbf{X})$. The condition number increases as more modes are included in the fitting procedure, and we discuss the impact of this phenomenon on our results in the following section. We do not impose any specific threshold on the condition number as a criterion for rejecting a model. All values of $\kappa$ reported here are well below the regime of numerical ill-conditioning. Here the condition number acts as a diagnostic: a larger $\kappa$ indicates stronger statistical correlations between the fitted multipole templates induced by the mask, which in turn inflates the variance of the recovered dipole amplitude and reduces its measured significance.

The results of this sequential fitting procedure are summarized in Table \ref{tab:lowl_dipole_results}.

\begin{table}[h!]
\centering
\begin{tabular}{lcccccc}
\toprule
 Model &
$D/M$ &
$l$ [deg] &
$b$ [deg] &
Offset from CMB [deg] &
Condition number ($\kappa$) \\
M + D &
0.01554 &
238.22 &
28.80 &
27.78 &
$5.05 \times 10^{0}$ \\

M + D + Q &
0.01545 &
237.86 &
29.18 &
27.68 &
$3.48 \times 10^{1}$ \\

M + D + Q + O &
0.01971 &
237.22 &
24.04 &
32.15 &
$5.66 \times 10^{1}$ \\

M + D + Q + O + $\ell{=}4$ &
0.01970 &
236.57 &
24.38 &
32.20 &
$2.28 \times 10^{2}$ \\
\bottomrule
\end{tabular}
\caption{Dipole estimates obtained from sequential low-multipole fits of the CatWISE2020 quasar map. The dipole amplitude is quoted relative to the monopole, $D/M$. The offset denotes the great-circle distance from the CMB dipole direction $(l,b) = (264.0^\circ, 48.3^\circ)$. The condition number refers to the normal-equation matrix $X^{\mathrm T}X$.}
\label{tab:lowl_dipole_results}
\end{table}

The dipole estimate remains stable when only the monopole and dipole are included. However, the inclusion of the octupole ($\ell=3$) shifts the amplitude by approximately $27\%$ and the direction by roughly 5$^\circ$, coinciding with an order of magnitude increase in the condition number. This indicates that the dipole estimate becomes sensitive to mask-induced mode coupling when higher multipoles are simultaneously fit, highlighting the importance of the multipole selection 
when fitting to the data. Consequently, the (M+D) fit, which has a low condition number, is adopted as our fiducial model for the \textsc{CatWISE2020} dipole.


\section{Significance estimates from different models}
\label{sec:a3}

The statistical significance of the dipole amplitude anomaly reported for the \textsc{CatWISE2020} quasar catalog is highly sensitive to the specific mathematical model used to extract the dipole from the sky map. While our primary analysis (Section~\ref{sec:sec6}) employed a simple monopole and dipole (M+D) fit, we systematically tested the impact of including higher multipoles in the fitting procedure.

\begin{figure}[htbp]
    \centering
    \begin{subfigure}[b]{0.48\textwidth}
        \centering
        \includegraphics[width=\textwidth]{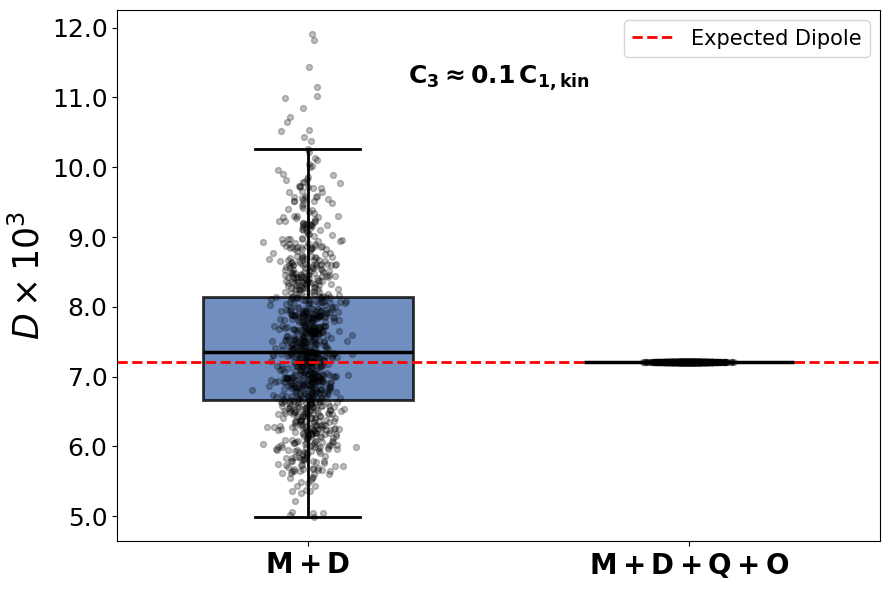}
        \label{fig:boxplot_c3}
    \end{subfigure}
    \hfill
    \begin{subfigure}[b]{0.48\textwidth}
        \centering
        \includegraphics[width=\textwidth]{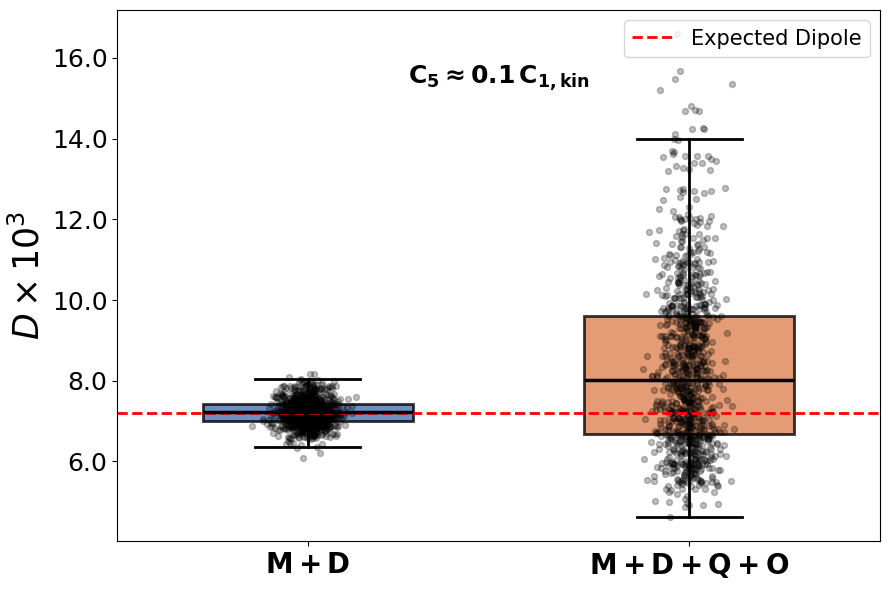}
        \label{fig:boxplot_c5}
    \end{subfigure}

    \caption{
    Recovered dipole amplitudes from simulations compared to the expected CMB dipole amplitude. 
    In both panels, the same sky mask is applied. 
    \textbf{Left panel:} The map includes only an test input $C_3 \approx 0.1 C_{1,kin}$ (octopole) component added to the kinematic dipole. 
    \textbf{Right panel:} The map includes only an test input $C_5\approx 0.1 C_{1,kin}$ component added to the kinematic dipole. 
    Each panel shows box plots comparing the $\mathrm{M+D}$ model (fitting only monopole and dipole) 
    with the $\mathrm{M+D+Q+O}$ model (fitting up to octopole). 
    This comparison demonstrates how including higher multipoles in the fit affects the recovery of the dipole amplitude.}
    \label{fig:boxplots_C3_C5}
\end{figure}

This sensitivity is particularly pronounced for the M+D+Q+O model. This choice, characterized by a high condition number (Appendix~\ref{sec:a2}), arises from strong mask-induced coupling between the dipole and higher-order multipoles particularly $\ell = 5, 7$ as shown in case of $\ell = 5$ in Figure \ref{fig:boxplots_C3_C5}, and beyond. Consequently, stochastic variations in these higher-$\ell$ modes within the simulated maps produce amplified variance in the estimated dipole amplitude. This inflated variance in the null distribution dilutes the statistical extremity of the observed \textsc{CatWISE2020} dipole, thereby suppressing its measured significance.

We recomputed the $p$-values for the \textbf{S1}, \textbf{S2}, and \textbf{S3} simulation suites using two alternative models: M+D+Q  and  M+D+Q+O. The results are presented in Table. \ref{tab:estimator_significance}

\begin{table}[h!]
\centering
\resizebox{\textwidth}{!}{%
\begin{tabular}{clccc}
\hline
\textbf{Simulation Suite} & \textbf{Model} & $n$ & $p$-value & \textbf{Significance ($\sigma$)} \\
\hline
\multirow{3}{*}{\textbf{S1} (No Clust. Dipole)}
 & M+D ($\kappa = 5.01$) & 13 & $1.30 \times 10^{-4}$ & 3.63 \\
 & M+D+Q ($\kappa = 34.8$) & 20 & $2.00 \times 10^{-4}$ & 3.53 \\
 & M+D+Q+O ($\kappa = 56.6$) & 797 & $7.97 \times 10^{-3}$ & 2.41 \\
\hline
\multirow{3}{*}{\textbf{S2} (Random Clust. Dipole)}
 & M+D ($\kappa = 5.01$) & 27 & $2.70 \times 10^{-4}$ & 3.44 \\
 & M+D+Q ($\kappa = 34.8$) & 32 & $3.20 \times 10^{-4}$ & 3.41 \\
 & M+D+Q+O ($\kappa = 56.6$) & 817 & $8.17 \times 10^{-3}$ & 2.40 \\
\hline
\multirow{3}{*}{\textbf{S3} (Aligned Clust. Dipole)}
 & M+D ($\kappa = 5.01$) & 58 & $5.80 \times 10^{-4}$ & 3.27 \\
 & M+D+Q ($\kappa = 34.8$) & 71 & $7.10 \times 10^{-4}$ & 3.19 \\
 & M+D+Q+O ($\kappa = 56.6$) & 1125 & $1.13 \times 10^{-2}$ & 2.28 \\
\hline
\end{tabular}%
}
\caption{Impact of model choice on the significance of the CatWISE2020 dipole excess. The reported $p$-value and significance ($\sigma$) are shown for three simulation suites (\textbf{S1}--\textbf{S3}) using the M+D (primary), M+D+Q, and M+D+Q+O models. $n$ denotes the number of simulations (out of 100,000) with a dipole amplitude $\geq$ the observed value.}
\label{tab:estimator_significance}
\end{table}

The results reveal a critical systematic: \textbf{the choice of   included in the fit dramatically alters the inferred significance.} The simple M+D and the M+D+Q models yield consistent results, with significances between $\sim 3.2\sigma$ and $\sim 3.6\sigma$ across the physical scenarios. In contrast, the M+D+Q+O model reduces the significance to approximately $2.3\sigma$--$2.4\sigma$, a difference of approximately $1\sigma$.

This reduction stems from a fundamental \textbf{bias–variance tradeoff} inherent to fitting multipoles to masked sky data. Including higher multipoles like the octopole in the fit reduces the bias from that specific multipole, but simultaneously increases the variance of the recovered dipole due to enhanced mask-induced effects. In particular when the M+D+Q+O model is applied to simulations containing realistic higher-order clustering modes (\textbf{S1}--\textbf{S3}), it yields a broader distribution of recovered dipole amplitudes. Consequently, the observed \textsc{CatWISE2020} amplitude, while large, becomes a less extreme outlier given the inflated variance, leading to a substantially higher $p$-value. Future work should therefore focus on systematic stress tests of dipole estimators under controlled conditions, including varying mask geometries, noise levels, and injected higher-order multipoles, in order to identify choices that provide the most reliable balance between bias and variance on a partial sky. 


\bibliography{references}{}
\bibliographystyle{aasjournalv7}

\end{document}